\documentstyle[12pt]{article}
\newcommand{\beq}{\begin{equation}}
\newcommand{\eeq}{\end{equation}}
\newcommand{\beqa}{\begin{eqnarray}}
\newcommand{\eeqa}{\end{eqnarray}}
\newcommand{\ba}{\begin{array}}
\newcommand{\ea}{\end{array}}
\newcommand{\CR}{\nonumber \\}
\newcommand{\pa}{\partial}
\newcommand{\A}{\alpha}

\newcommand{\D}{\delta}          

\newcommand{\E}{\epsilon}
\newcommand{\La}{\Lambda}
\newcommand{\p}{\Phi}
\newcommand{\bra}{\langle}
\newcommand{\ket}{\rangle}

\newcommand{\tr}{{\rm Tr}}

\newcommand{\CP}{{\bf C}{\rm P}}

\newcommand{\Z}{{\bf Z}}

\newcommand{\pare}[1]{{\left( #1 \right)}}

\newcommand{\sq}[1]{{ #1^{\frac{1}{2}} }}

\begin{document}

\begin{titlepage}
\null
\begin{flushright} 
hep-th/9712172  \\
UTHEP-377 \\
December, 1997
\end{flushright}
\vspace{0.5cm} 
\begin{center}
{\Large \bf
Supersymmetric Gauge Theories 

with Classical Groups via M Theory Fivebrane
\par}
\lineskip .75em
\vskip2.5cm
\normalsize
{\large Seiji Terashima} 
\vskip 1.5em
{\it Institute of Physics, University of Tsukuba, Ibaraki 305, Japan}
\vskip3cm
{\bf Abstract}
\end{center} \par
We study the moduli space of vacua of 
four dimensional $N=1$ and $N=2$ supersymmetric gauge theories with 
the gauge groups $Sp(2 N_c)$, $SO(2 N_c)$ and $SO(2 N_c +1)$
using the M theory fivebrane.
Higgs branches of the $N=2$ supersymmetric gauge theories are
interpreted in terms of the M theory fivebrane
and the type IIA $s$-rule is realized in it.
In particular we construct the fivebrane configuration 
which corresponds to a special Higgs branch root.
This root is analogous to the baryonic branch root in the $SU(N_c)$ theory 
which remains as a vacuum after the adjoint mass perturbation 
to break $N=2$ to $N=1$.
Furthermore we obtain the monopole condensations and 
the meson vacuum expectation values 
in the confining phase of $N=1$ supersymmetric gauge theories
using the fivebrane technique.
These are in complete agreement with the field theory results
for the vacua in the phase with a single confined photon.

\end{titlepage}

\baselineskip=0.7cm

\section{Introduction}

Recently the D(irichlet)-brane 
in flat spacetime
has provided very interesting tools to understand
strong coupling dynamics of gauge theory in various dimensions.
This approach was initiated in the work of Hanany and Witten 
in type IIB setup \cite{HaWi}.

The four dimensional gauge theory is realized 
on the worldvolume of D branes which are suspended between
the two NS 5-branes in type IIA string theory \cite{ElGiKu} \cite{ElGiKuRaSc}.
It is shown that one can give an exact
low energy description of the $N=2$ $SU(N_c)$ gauge theory 
by reinterpreting the brane configuration from M theory point of view
where the D4-branes and NS 5-branes of the type IIA theory 
are unified into the fivebrane of M theory \cite{Wi}.
In this framework, the Higgs phase of 
$N=2$ $SU(N_c)$ gauge theory is investigated intensively 
\cite{HoOoOz} \cite{NaOhYoYo}.

The $N=1$ perturbation by the mass term of $\p$,
which is the scalar chiral multiplet 
in the adjoint representation of the gauge group,
induces the rotation of $N=2$ fivebrane configuration \cite{Ba}.
In this way one can study the strong coupling dynamics of 
$N=1$ $SU(N_c)$ gauge theory 
\cite{HoOoOz} \cite{Wi2}-\cite{25}.
Furthermore one can construct the fivebrane configuration 
that describes the $N=1$ $SU(N_c)$ gauge theory 
with the superpotential $\Delta W=\sum \mu_{2 k} \tr \p^{2 k}/ 2k$ by
imposing the appropriate boundary conditions on the fivebrane configuration 
\cite{DeOz}.
The fivebrane contains the complete informations
of the the monopole condensations and 
the meson vacuum expectation values (vev)
in the confining phase of corresponding $N=1$ supersymmetric gauge theory.

In this article 
we first study the Higgs phase of $N=2$ gauge theory with 
the classical gauge groups in the fivebrane framework.
To carry out this, we introduce the orientifold 
in M theory fivebrane configuration
\cite{ElGiKuRaSc} \cite{EvJoSh}-\cite{BrSoThYa}.
We find the agreement between the field theory results of 
the Higgs phase of $N=2$ gauge theory \cite{ArPlSh2} \cite{HiMaSu}
and the results from the fivebrane picture.
Next we extend the analysis of \cite{DeOz} to the theory with 
the classical gauge groups.
The monopole and meson vevs are computed
using the fivebrane configuration 
in the vacua where photons are confined.
We again find the agreement with the field theory results.

Section 2 is devoted to field theory analysis of 
the theory with the gauge group $Sp(2 N_c)$.
In particular we study the monopole and meson vevs
of the $N=1$ theory obtained by adding to the $N=2$ superpotential
an $N=1$ perturbation of the form $\Delta W=\sum \mu_{2 k} \tr \p^{2 k}/ 2k$.
In section 3, we construct the fivebrane configuration 
describing the Higgs branches of the 
$N=2$ supersymmetric QCD with the gauge group $Sp(2 N_c)$.
In section 4, we rotate the $N=2$ fivebrane configuration 
corresponding to the adjoint mass perturbation 
in such a way that the resulting fivebrane configuration 
describes the $N=1$ theory with the superpotential $\Delta W$.
Then the fivebrane results are compared 
to those obtained in section 2.
In sections 5 and 6, we consider the $SO(2 N_c)$ and $SO(2 N_c+1)$ 
gauge theories respectively.
Finally we draw our conclusions.

\section{Field theory analysis of $Sp(2 N_c)$ gauge theory}\label{field}

Following the paper of \cite{HoOoOz} and \cite{DeOz},
we consider the case of the gauge group $Sp(2 N_c)$ in section 2,3 and 4.
In this section the field theory analysis of 
the theory with the gauge group $Sp(2 N_c)$ is performed.
The results obtained in this section will be compared to the 
results from the point of view of M theory.

\subsection{$N=2$ Moduli space of vacua}

Let us consider $N=2$ supersymmetric gauge theory
with the gauge group $Sp(2 N_c)$ and $N_f$ quark hypermultiplets 
in the fundamental representation $Q_a^i$, $i=1,\ldots, 2 N_f$.
In term of the $N=1$ superfields,
the vector multiplet consists of a field strength chiral multiplet 
${(W^\A)}^{\;\;a}_{b}$ and a scalar chiral multiplet $\Phi_{a}^{\;\;b}$ 
both in the adjoint representation of the gauge group.
Here $a,b=1, \ldots,2 N_c$ are color indices.
The $2N_c \times 2N_c$ tensor $\p$ is subject 
to ${}^t \Phi = J \Phi J $ with
the symplectic metric $J={\rm diag}(i\sigma_2, \cdots, i\sigma_2)$
where $\sigma_2=\pmatrix{0 & -i \cr i & 0}$.
The $N=2$ superpotential takes the form 
\beq
W=\sqrt{2} Q_a^i J^{ab} \Phi_{b}^{\;\; c} Q_{c}^{i} +\sqrt{2} 
m^{ij} Q_a^i J^{ab} Q_b^j,
\label{W}
\eeq
where $m$ is quark mass matrix. From the $N=2$ supersymmetry we can take
$m={\rm diag}(i m_1 \sigma_2, \cdots, i m_{N_f} \sigma_2)$.

If the quark mass $m$ vanishes, 
the classical global symmetry groups are 
the flavor symmetry $O(2 N_f)=SO(2 N_f) \times \Z_2$ 
in addition to $SU(2)_R \times U(1)_R$ R-symmetry group.
The theory is asymptotically free for $N_f<2 N_c+2$.
The $U(1)_R$ symmetry is anomalous and is broken down 
to $\Z_{2 N_c+2-N_f}$ 
since the instanton factor is proportional to $\La^{2 N_c+2-N_f}_{N=2}$ 
where $\La_{N=2}$ is the dynamically generated scale.
Note that $\Z_2$ of the flavor symmetry is also broken down
because the instanton factor changes its sign by the action of the $\Z_2$.

The moduli space of the vacua consists of the Coulomb and Higgs branches.
The Coulomb branch is $N_c$ complex dimensional and 
is parametrized by the gauge invariant order parameters 
\beq
u_{2 k}= \left\langle \frac{1}{2 k} {\rm Tr} \left( \p^{2k} \right) 
\right\rangle, \;\;\; k=1, \ldots ,N_c .
\label{gi}
\eeq
The Coulomb branch parametrize a family of 
genus $N_c$ hyperelliptic curves \cite{ArSh}
\beq
x y^2=\left( x B_{2 N_c}(x,u_{2 k}) +
\La^{2 N_c+2-N_f}_{N=2} i^{N_f} \prod_{i=1}^{N_f} m_i \right)^2
-\La^{2(2 N_c+2-N_f)}_{N=2} \prod_{i=1}^{N_f} (x-m_i^2),
\label{spcurve}
\eeq
where $B_{2 N_c}$ is a degree $N_c$ polynomial in $x$ with coefficients 
that depend on the gauge invariant order parameters $u_{2 k}$ 
and $m_i$.
For $N_f < N_c+2$ the polynomial $B_{2 N_c}$ is given by
\beq
B_{2 N_c}(x)=\sum_{i=0}^{N_c} s_{2 i} x^{N_c-i} ,
\eeq
where $s_{2 k}$ and $u_{2 k}$ are related by the Newton formula
\beq
k s_{2 k} + \sum_{j=1}^k j s_{2(k-j)} u_{2 j} =0, \;\;\; k=1,2, \ldots ,N_c,
\eeq
where $s_0=1$.
We also define $s_{2 k}=0$ for $k>N_c$, $k<0$.
From these we can see that the relation
\beq
\frac{\pa s_{2 j}}{\pa u_{2 k}} = -s_{2(j-k)},
\label{sp1}
\eeq
holds.

In the rest of this section we will consider the case with 
the vanishing bare quark mass.
There is only one type of the gauge invariants which are constructed from 
$Q$, the meson fields $M^{ij}=Q^i J Q^j$.
The Higgs branches are classified by an integer $r$ such that $0 \leq r \leq {\rm min} \{ N_c,\frac{N_f}{2} \} $ \cite{ArPlSh2}.
The $r$-th Higgs branch has complex dimension $2 r (2 N_f-2 r-1)$
and the root of this branch is its $N_c-r$ dimensional 
submanifold of intersection with 
the Coulomb branch.
The effective theory along the root of the $r$-th Higgs branch
is $Sp(2 r) \times U(1)^{N_c-r}$ with $N_f$ massless quarks 
which are neutral with respect to any $U(1)$ factor.
The curve at the $r$-th Higgs branch root takes the form
\beq
y^2 = x^{2r+1  }\left( 
B_{2(N_c-r)}(x)^2  - \Lambda_{N=2}^{2(2N_c+2-N_f)} x^{N_f-2r-2} \right ),
\label{spcurvenb}
\eeq
where $B_{2( N_c-r)}$ is a degree $(N_c-r)$ polynomial in $x$.
At the special points along the root,
additional massless hypermultiplets charged under the $U(1)$'s exist.
In particular in the $r^*$-th Higgs branch where $r^*=N_f-N_c-2$,
the maximal singularity occurs at which the curve takes the form
\beq
y^2=x^{2 r^*+1} \left( x^{N_c-r^*}+\frac{1}{4} \La^{2(N_c-r^*)} \right)^2,
\label{spmax}
\eeq
and $N_c-r^*$ hypermultiplets become simultaneously massless.

\subsection{Breaking $N=2$ to $N=1$}

Next we consider the perturbation $\Delta W$ to the 
$N=2$ superpotential (\ref{W}) 
\beq
W=\sqrt{2} Q J \Phi Q +\sqrt{2} m Q J Q+ \Delta W,
\label{Wp}
\eeq
where 
\beq
\Delta W=\sum_{k=1}^{N_c} 
\mu_{2 k} \frac{1}{2 k} {\rm Tr} (\p^{2 k}),
\label{sppert}
\eeq 
which break the $N=2$ supersymmetry to the $N=1$ supersymmetry.
If only the mass perturbation,
say $\Delta W=\mu_2 \frac{1}{2} \tr \p^2 $, is present,
only the points in the $r^*$-th Higgs branch root 
at which the curve takes the form (\ref{spmax}) are not lifted \cite{ArPlSh2}.
Moreover even in $N_f< N_c+2$ case the point in the Coulomb branch
where the curve degenerates
to the genus zero curve 
such as $y^2=x^{2 n} \prod_{i=1}^{N_c-n} (x-a_i)^2 (x-b)$ where $n$ is some integer
is also expected to remain as vacua.
Because if we take the limit $m_i \rightarrow \infty$ for all $i$
the theory becomes the pure Yang-Mills theory and
the $N=2$ curve degenerates to the genus zero curve 
at some discrete points in the moduli space of the pure Yang-Mills theory.

Now we consider $N=2$ pure $Sp(2 N_c)$ Yang-Mills theory
perturbed by the superpotential $\Delta W$ (\ref{sppert}).
Near points where $N_c$ or less mutually local dyons are massless, 
the low energy effective superpotential is
\beq
W =  \sqrt{2} \sum_{i=1}^{N_c} \tilde{M}_i A_i M_i 
+ \sum_{k=1}^{N_c}\mu_{2 k} U_{2 k},
\label{spWp}
\eeq
where $A_i$ are the $N=1$ chiral superfields in the $N_c$ $N=2$ $U(1)$ vector
multiplets, $M_i,\tilde{M}_i$ are the dyon hypermultiplets and
$U_{2 k}$ represent the superfields corresponding to Casimirs 
$\frac{1}{2 k} \tr (\Phi^{2 k})$ \cite{SeWi} \cite{ArDo}.
We will use lower-case letters to denote the lowest components of the 
corresponding upper-case superfields.

As in the $SU(N_c)$ case \cite{DeOz}, 
the holomorphy and global symmetries 
guarantee that the superpotential (\ref{spWp}) is exact.
The list of charges of  
the parameters and fields under $U(1)_R$ and $U(1)_J \subset SU(2)_R$ 
is given by  
\beq
\begin{array}{cccl}
& U(1)_R & U(1)_J & \\
A_i & 2 & 0 & \\
\tilde{M}_i M_i & 0 & 2 & \\
\mu_{2 k} & 2 - 4 k & 2 & \\
U_{2 k} & 4 k & 0 & \\
\La_{N=2} & 2 & 0 &
\end{array}
\label{list1}
\eeq

We find the vacua of the low energy theory by imposing 
the vanishing D-term constraints $|m_i| = |\tilde{m}_i| $
and the equation of motion
\beq
-\frac{\mu_{2 k}}{\sqrt{2}} = 
\sum_{i=1}^{N_c}\frac{\pa a_i}{\pa u_{2 k}} m_i \tilde{m}_i, 
\;\;\;\;\; k=1, \ldots, N_c,
\label{vac1} 
\eeq
and 
\beq
a_i m_i = a_i \tilde{m}_i = 0, \;\;\;\;\; i=1,\ldots, N_c \;\; .
\label{vac} 
\eeq
This implies that the points of the moduli space of the vacua 
are lifted by the perturbation if there are no massless dyons.

Next we consider a point in the moduli space where 
$l$ mutually local dyons
are massless. This means that some of the one-cycles shrink to zero and
that the genus $N_c$ curve (\ref{spcurve}) degenerates 
to a genus $N_c-l$ curve.
Thus the curve (\ref{spcurve}) takes the form
\beq
y^2=x { B_{2 N_c}(x) }^2 + 2 B_{2 N_c}(x) \La_{N=2}^{2 N_c+2} =
\prod_{i=1}^{l}(x-p_i^2)^2
\prod_{j=1}^{2 N_c+1-2l} (x-q_j^2)
\label{sppr}
\eeq
with $p_i$ and $q_j$ distinct.
The equation of motion implies that
$ m_i=\tilde{m}_i=0$ for $i = l+1, \ldots ,N_c$
while $m_i, \tilde{m}_i$ for
$i=1, \ldots, l$ are unconstrained. Since the matrix
$\partial a_i / \partial u_{2 k}$ is non-degenerate (see (\ref{spau22}))
and there are $N_c-l$ free parameters
in the form (\ref{sppr}), there will be a 
complex $N_c-l$  dimensional moduli space of $N=1$ vacua which remains
after the perturbation.

The matrix $\partial a_i/\partial u_{2 k}$
can be explicitly evaluated using the relation
\beq
\frac{\partial a_i}{\partial s_{2 k}}
= \oint_{\beta_i} \frac{x^{N_c-k} dx}{y} ,
\label{spau20}
\eeq
where the rhs is the integral of a  holomorphic one form on the
curve (\ref{spcurve}). 
At a point
where the $l$ dyons become massless the cycles
$\beta_i \rightarrow 0$ $(i=1 \ldots l)$ and
(\ref{spau20}) becomes to a contour integrals around $x=p_i^2$ 
where $ i=1, \ldots, l$.
Then we obtain 
\beq
 \label{spau22}
 \frac{\partial a_i}{\partial s_{2 k}} = \frac{p_i^{2(N_c-k)}}
 { \prod_{l \neq i}(p_i^2-p_l^2) \prod_m (p_i^2-q_m^2)^{1/2} }.
\eeq
From  (\ref{sp1}),(\ref{vac1})  
and (\ref{spau22}) we find the relation between 
the parameters $\mu_{2 k}$ and the dyon vevs $m_i \tilde{m}_i$ as
\beq
 \label{spau23}
\frac{-\mu_{2 k}}{\sqrt{2}} = 
\sum_{i=1}^l \sum_{j=1}^{N_c} (- s_{2(j-k)}) p_i^{2(N_c-j)}
\frac{m_i \tilde{m}_i}{ 
\prod_{l \neq i}(p_i^2-p_l^2) \prod_m (p_i^2-q_m^2)^{1/2} }.
\eeq
For later convenience we define
\beq
\omega_i = \frac{\sqrt{2} m_i \tilde{m}_i}{
\prod_{l \neq i}(p_i^2-p_l^2) \prod_m (p_i^2-q_m^2)^{1/2} },
\label{spwi}
\eeq
and rewrite the generating function 
$\sum_{k=1}^{N_c} \mu_{2 k} v^{2 k-1}$ for the $\mu_{2 k}$ as the form
\beqa
\sum_{k=1}^{N_c} \mu_{2 k} v^{2 k-1} & = & \sum_{k=1}^{N_c} \sum_{i=1}^l 
\sum_{j=1}^{N_c} v^{2 k-1} s_{2(j-k)} p_i^{2(N_c-j)} \omega_i \nonumber\\{}
& = &  \sum_{i=1}^l 
\sum_{j=1}^{N_c} B_{2 N_c}(v^2) v^{2(j-N_c)-1} p_i^{2(N_c-j)} \omega_i 
  + {\cal O}(v^{-1}) \nonumber\\{}
& = &  v B_{2 N_c}(v^2) \sum_{i=1}^l  \frac{\omega_i}{(v^2-p_i^2)} 
  + {\cal O}(v^{-1}),
\label{spau30}
\eeqa
where we have set $x$ as $v^2$.

When perturbation parameters $\mu_{2 k}$ are given
and a point in the $N=2$ 
moduli space of vacua is specified by the set $p_i,q_j$ of (\ref{sppr}),
the equation (\ref{spau30}) determines whether this point remains 
as an $N=1$ vacuum after the perturbation
and fixes the vevs of the dyon fields $m_{i} \tilde{m}_i$.
This is so since on the lhs of (\ref{spau30}) 
we have $N_c$ couplings $\mu_{2 k}$ $(k=1,\ldots, N_c)$,
while on the rhs there are $l$ dyon condensates $\omega_i$
and $(N_c-l)$ independent parameters among 
$p_i$'s and $q_j$'s as is seen from (\ref{sppr}).

We remark here that in the perturbation $\Delta W$ (\ref{sppert})
we choose a particular basis of Casimirs $u_{2 k}$.
It may be allowed to take 
different basis $\tilde{u}_{2 k}=\tilde{u}_{2 k}(u_{2 i})$.
In fact this is an useful ambiguity as we will observe when 
discussing the meson vev in view of the confining phase and 
the fivebrane in M theory.
If we adopt the perturbation 
$\Delta W = \sum_{k=1}^{N_c}\mu_{2 k} \tilde{u}_{2 k}$,
We define $\tilde{\mu}_{2 k}$ from the equation
\beq
\frac{\pa \Delta W}{\pa \p} = 
\sum_{k=1}^{N_c} \mu_{2 k} \sum_{n=1}^{N_c} \p^{2n -1} 
\frac{\pa \tilde{u}_{2 k}}{\pa u_{2 n}}
=\sum_{n=1}^{N_c} \tilde{\mu}_{2 n} \p^{2n -1} .
\eeq
Thus as in the previous considerations, 
the dyon condensation is expressed as
\beq
-\frac{\mu_{2k}}{\sqrt{2}} = 
\sum_{i=1}^{N_c}\frac{\pa a_i}{\pa \tilde{u}_{2 k}} m_i \tilde{m}_i, 
\;\;\;\;\; k=1, \ldots, N_c.
\label{vac1a} 
\eeq
And if we define $w_i$ as Eq. (\ref{spwi}), we obtain that
\beq
\mu_{2k} = -\sum_{i=1}^{l} \sum_{j=1}^{N_c} 
\frac{\pa s_{2 j}}{\pa \tilde{u}_{2 k}} p_i^{2(N_c-j)} w_i
= \sum_{i=1}^{l} \sum_{j=1}^{N_c} \sum_{m=1}^{N_c} 
s_{2(j-m)} \frac{\pa u_{2 m}}{\pa \tilde{u}_{2 k}} p_i^{2(N_c-j)} w_i.
\label{vac1b} 
\eeq
Therefore we find that 
\beqa
\sum_{n=1}^{N_c} \tilde{\mu}_{2n } v^{2 n-1} & =& 
\sum_{k=1}^{N_c} \sum_{n=1}^{N_c} 
\mu_{2 k} \frac{\pa \tilde{u_{2 k}}}{\pa u_{2 n}} v^{2n-1}
=\sum_{i=1}^{l} \sum_{j=1}^{N_c} \sum_{n=1}^{N_c}
s_{2(j-n)} v^{2 n-1} p_i^{2(N_c-j)} w_i \CR
&=& v B_{2 N_c}(v^2) \sum_{i=1}^l  \frac{\omega_i}{(v^2-p_i^2)} 
  + {\cal O}(v^{-1}),
\label{vac1c} 
\eeqa
which is the same form as (\ref{spau30}).

There are $N_c+1$ points in the moduli space related 
to each other by 
the action of the discrete $\Z_{2 N_c+2}$
$R$-symmetry group (\ref{list1}),
where $N_c$ mutually local dyons are massless.
At these points $a_i=0,  i=1,...,N_c$ and the curve (\ref{spcurve})
degenerates to a genus zero curve. 
These points correspond to the 
$N_c+1$ massive vacua of  $N=1$ pure Yang-Mills theory
where the discrete $\Z_{2 N_c+2}$
$R$-symmetry is spontaneously broken to $\Z_2$.
Equations (\ref{vac1})
can be solved for generic $\mu_{2 k}$  and these $N=1$ vacua are 
generically not lifted. 


Now we introduce $N_f$ flavors 
in the $Sp(2 N_c)$ gauge theory.
As in the pure Yang-Mills case, the perturbation (\ref{sppert}) 
lifts the non singular locus of the
$N=2$ Coulomb branch. The computation of the dyon vevs 
along the singular locus which does not become 
Higgs branch root is similar to 
the pure Yang-Mills case.

In the following we will compute the dyon vevs at the 
roots of the Higgs branches.
The effective theory along the root of the $r$-th Higgs branch 
is $Sp(2 r) \times U(1)^{N_c-r}$
with $N_f$ massless quark multiplets $Q$.
We assume here that 
the superpotential describing $\tilde{M}_i,M_i$ 
and the vector multiplet $A_i$ for each $U(1)$ factor at these points is 
\beq
W = \sqrt{2}\sum_{i=1}^{N_c-r} \tilde{M}_i A_i M_i  
+ \sum_{k=1}^{N_c-r} \mu_k  U_k,
\label{WNB}
\eeq
where we set $\mu_{2 k}=0$ $(k \geq N_c-r+1)$ as in \cite{DeOz}.
Then from the equation of motion,
most of the Higgs branch roots are lifted unless some of
the hypermultiplets $\tilde{M}_i,M_i$ become massless.
Thus we will consider the point where
the hypermultiplets $\tilde{M}_i,M_i$ $(1 \leq i \leq l)$ are massless.
Taking the coordinate $\tilde{y}=\frac{y}{x^r}$, 
the curve (\ref{spcurvenb}) becomes the form
\beq
\tilde{y}^2=
x \left ( {B_{2(N_c-r)}(x)}^2-\Lambda_{N=2}^{2N_c+2-N_f} x^{N_f-2r-2} \right )
= \prod_{i=1}^{l}(x-p_i^2)^2 \prod_{j=1}^{2(N_c-r-l)+1}(x-q_j^2). 
\label{spyt}
\eeq
Repeat the analysis of the pure $Sp(2 N_c-2 r)$ Yang-Mills case 
with the relation
\beq
\frac{\partial a_i}{\partial s_{2 k}}
= \oint_{\beta_i} \frac{x^{N_c-k} dx}{y}
= \oint_{\beta_i} \frac{x^{N_c-k-r} dx}{\tilde{y}} ,
\label{spaa}
\eeq
then we obtain 
\beq
\sum_{k=1}^{N_c-r} \mu_{2 k} v^{2 k-1} 
 =   2 H(v^2) \frac{v B_{2 (N_c-r)}(v^2)}{\prod_{i=1}^l  (v^2-p_i^2)} 
  + {\cal O}(v^{-1}).
\label{spbb}
\eeq

\subsection{Meson vev}


We now  compute the vev of the meson field
$Q J Q$ along the singular locus of the Coulomb branch.
This vev is generated by the non perturbative dynamics of 
the $N=1$ theory, and was clearly
zero in the $N=2$ theory before the perturbation (\ref{sppert}).

Using the technique of the confining phase superpotentials
\cite{ElFoGiInRa}-\cite{TeYa2}, 
we can determine a low-energy effective superpotential for
the phase with a confined photon. From that, we determine the
meson vevs.
We take a tree-level superpotential \cite{KiTeYa}
\beq
W=\sum_{n=1}^{N_c-1} \mu_{2 n} u_{2 n} + \mu_{2 N_c} s_{2 N_c}
+\sqrt{2}  Q J \p Q
+\sqrt{2}  m Q J Q.
\label{wsp}
\eeq
Note here that we choose $s_{2 N_c}$ instead of $u_{2 N_c}$ 
as the top Casimir.
Due to this modification, the theory with 
the superpotential (\ref{wsp}) does not allow
vacua with an unbroken $Sp(2) \times U(1)^{N_c-1}$ gauge symmetry classically
and recovers the $N=2$ curve correctly.
Then we take the classical vacua with $Q=0$ and 
an unbroken $SU(2) \times U(1)^{N_c-1}$ gauge symmetry
which corresponds to the phase with a confined photon.
Including the contributions from 
the instanton effect in the broken part of the gauge group 
and the gaugino condensation 
in the low energy pure $SU(2)$ Yang-Mills theory,
we obtain the effective superpotential 
\beqa
W(\mu_{2 k},m_i)
& =& \sum_{n=1}^{N_c-1} \mu_{2 n} u_{2 n}^{cl} 
+ \mu_{2 N_c} s_{2 N_c}^{cl} \CR
& & +\frac{\mu_{2 N_c}}{p_1^2} \La_{N=2}^{2 N_c+2-2 N_f} 
\left( i^{N_f} \prod_{i=1}^{N_f} m_i \pm
\prod_{i=1}^{N_f} (p_1^2-m_i^2)^{\frac{1}{2}}  \right),
\label{wsp1}
\eeqa
where $p_1^2=\frac{\mu_{2 (N_c-1)}}{\mu_{2 N_c}}$
and the normalization of $\La_{N=2}$ is fixed by the
$N=2$ curve (\ref{spcurve}).
Note that $p_1$ is identical to the one
in the degenerate curve (\ref{sppr}) with $l=1$.
Here we assume that the effective superpotential (\ref{wsp1}) is exact
\cite{In} \cite{ElFoGiInRa}.
Defining $M^i$ as the $i$-th eigenvalue of the meson field $M$
and using 
$\sqrt{2} \bra M^{i} \ket = \pa W(\mu_{2 k},m_i) / \pa m_i$,
we find the meson vev
\beq
\bra M^{i} \ket = \frac{\mu_{2 N_c}}{\sqrt{2} p_1^2 } 
\La_{N=2}^{2 N_c+2-2 N_f} 
\left( \frac{ i^{N_f} \prod_{k=1}^{N_f} m_k}{m_i}  \mp
m_i \frac{ \prod_{i=1}^{N_f} (p_1^2-m_i^2)^{\frac{1}{2}} }{p_1^2-m_i^2}  \right).
\label{spbb8}
\eeq

This method seems to apply for the phase with more confined photons.
But for example in the case of the gauge group $SU(4)$ broken to 
$SU(3) \times U(1)$ or $SU(2) \times SU(2) \times U(1)$ classically,
the low-energy effective superpotential obtained in this method 
can not derive the singular locus of the $N=2$ theory.
This can be seen from the explicit calculation.
Furthermore we see that the meson vev computed from 
this low-energy effective superpotential
is not correct for the case of the gauge group $SU(N_c)$ \cite{DeOz}.
Thus for the phase with more confined photons,
we expect this method is not reliable 
and we will not discuss the meson vev from field theory analysis.
Later this will be computed from the fivebrane configuration in M theory.

\section{$N=2$ Higgs branch of $Sp(2 N_c)$ theory and M theory}

In this section we study the moduli space of vacua of 
$N=2$ supersymmetric QCD with the gauge group $Sp(2 N_c)$ and
its deformations by the superpotential (\ref{sppert})
by using the fivebranes in M theory.

Let us consider first the type IIA string theory on a flat space-time.
Following the paper of \cite{LaLoLo},
we consider the type IIA picture of the $N=2$ gauge theory 
with the gauge group $Sp(2 N_c)$.
Consider the brane configuration which consists of
two NS 5-branes with worldvolume
coordinates $x^0,x^1,x^2,x^3,x^4,x^5$
at $x^7=x^8=x^9=0$,
$N_c$ D4-branes
suspended between them
with  worldvolume
coordinates $x^0,x^1,x^2,x^3,x^6$ at $x^7=x^8=x^9=0$
and $N_f$ D6-branes with 
worldvolume
coordinates $x^0,x^1,x^2,x^3,x^7,x^8,x^9$.
In addition to this we introduce 
an orientifold four plane parallel to the D4-branes.
Corresponds to this, we mod out the spacetime by the reflection
$(x^4,x^5,x^7,x^8,x^9) \rightarrow (-x^4,-x^5,-x^7,-x^8,-x^9)$,
together with the gauging of world sheet parity $\Omega$.
In order to obtain the $Sp(2 N_c)$ gauge group,
the Chan-Paton wavefunction of the vector is taken to be symmetric.
Each object which is not invariant under the reflection induced by the 
orientifold plane has a mirror partner.
We take the configuration in which 
the all D6-branes have their mirror partners but
the NS 5-branes do not have mirror partners
since we will not move the NS 5-branes.
Another important aspect of the orientifold is its RR charge. In the normalization, 
where a D4-brane and its mirror partner carries one unit of this charge, 
the charge of the orientifold plane is $1$ in the D4-brane sector.
Note that
if we take the $SO$ gauge group, the charge of the orientifold plane is $-1$.

Since the D4-branes are finite in the $x^6$ direction,
the effective world volume theory of the D4-branes becomes
the four dimensional $N=2$ supersymmetric QCD with the gauge group $Sp(2 N_c)$
and $N_f$ flavors. 
The classical $U(1)_R$ and $SU(2)_R$
R-symmetry groups of the four-dimensional theory
on the D4-branes worldvolume are interpreted as 
the rotations in the $x^4,x^5$ and $x^7,x^8,x^9$ directions.

The Coulomb branch of the theory is parametrized by the motions of
D4-branes along the NS 5-branes.
If some D4-branes lie on two or more D6-branes,
we find that the phase of the theory can shift to the Higgs branch
to break the D4-branes on the D6-branes
and have them suspended between the D6-branes.
Motions of the D4-branes along the D6-branes 
describe the Higgs branch.
The location of a D4-brane between two D6-branes is parametrized 
by two complex parameters,
the $x^7,x^8,x^9$ 
coordinates  together with the gauge field component  $A_6$
in the $x^6$ coordinate. 

The $r$-th Higgs branch 
corresponds to $2 (N_c - r)$  D4-branes suspended between
the two NS 5-branes and $2 r$ D4-branes
broken on the D6-branes.
Since the $s$-rule \cite{HaWi} 
does not allow more than one D4-brane
to be suspended between a NS 5-brane
and a D6-brane, $r$ can not be greater than $[N_f/2]$.
Since the Coulomb branch in the brane picture corresponds to  
D4-branes moving along the two NS 5-branes,
the $r$-th Higgs branch shares $(N_c-r)$ complex
dimensions with the Coulomb branch,
corresponding to the the gauge group $Sp(2(N_c-r))$.

The complex dimension of the $r$-th Higgs
branch in the Higgs direction is determined
by counting the number of the D4-branes suspended 
between  the D6-branes \cite{ElGiKuRaSc}. 
If we ignore the effect of the orientifold,
the configuration with the $2 r$ D4-branes broken
on the D6-branes
is identical with the one appearing in the $2 r$-th non-baryonic branch of the 
$SU(2 N_c)$ gauge theory with $2 N_f$ flavors.
In this configuration 
the $j$-th D4-brane ($1 \leq j \leq 2 r$) is broken 
on the $i$-th D6-branes ($j \leq i \leq 2 N_f+1-j$)
and the $k$-th D4-brane ($2r+1 \leq k \leq 2 N_f$) is not broken.
Here the $(2 n-1)$-th and the $2n$-th D6-branes
are interchanged under the reflection due to the orientifold.
Now taking into account the effect of the orientifold,
the components of the broken D4-branes must be paired.
Then in this configuration the components of the D4-branes which 
are not paired are fixed on the orientifold and 
the components of the $j$-th pair of D4-branes ($1 \leq j \leq r$) are suspended 
between the $i$-th D6-branes where $2 j \leq i \leq 2 N_f+1-2 j$.
Thus we obtain the dimension to be 
\beq
2 \sum_{l=1}^r \left[ 2 N_f-(4l-3)-2 \right] = 2r(2 N_f-2 r-1)
\label{spdH}
\eeq
in agreement with the field theory results.

Now we consider the effect of the intersection of branes and orientifold
as in \cite{GiKu}.
The charge of the orientifold is $+1$ in the part between the 
two NS 5-branes and it is $-1$ in the part outside the NS 5-branes.
Following \cite{deHoOoOz}, we expect that
between the two NS 5-branes
there are two D4-branes whose degrees of freedom 
corresponding to the motion away from the orientifold are frozen.
Furthermore 
these fixed D4-branes are expected to be 
absent between the mirror pair of the D6-branes.
On the other hand in the case of the $SO(2 N_c)$ gauge theory,
the charge of the orientifold is $-1$ in the part between the 
two NS 5-branes and it is $+1$ in the part outside the NS 5-branes.
Thus we expect that outside the NS 5-branes and 
between the mirror pair of the D6-branes,
there are two fixed D4-branes.

In the above consideration of 
counting the dimension of the Higgs branch,
we have thought that 
the components of the D4-branes between the $k$-th D6-brane and
the $k+1$-th D6-brane are paired under the reflection.
If $k$ is odd there remains a component of the D4-brane
which is not paired. 
This remaining component has no degrees of freedom.
However if fixed D4-branes exist 
we suppose that this remaining component of the D4-brane
can be paired with the fixed D4-brane and may have the degrees of freedom.
Although we have no convincing arguments about how this phenomenon occurs,
we will assume this hereafter and see the validity of this assumption.
In the case of the $Sp(2 N_c)$ gauge theory,
this assumption does not alter the above consideration
of counting the dimension of the Higgs branch.
However in the case of the $SO(1)$ gauge theory,
we easily see by virtue of our assumption 
that the dimension of the Higgs branch is $2 N_f$.
This coincides with the dimension of 
the moduli space of vacua of the theory of $2 N_f$
free chiral superfields which is equivalent to 
the $SO(1)$ gauge theory.
Later we will see that 
the correct dimensions of the Higgs branches are obtained 
in the case of the $SO(n)$ gauge theory $(n>1)$ under our assumption.
This consideration can also apply to counting 
the dimension of the Higgs phase of the 
$N=1$ supersymmetric QCD with classical groups.
The results are in agreement with the 
field theory results.

In \cite{AhOhTa} and \cite{AhOhTa2} the points in the 
moduli space where all the eigenvalues of meson vev are equal
after the $N=1$ perturbation are interpreted as
belonging to the $r$-th Higgs branch with $r=N_c$.
In the light of the $SU(N_c)$ result based on the global symmetry \cite{HoOoOz}
and the field theory result of the $Sp(2 N_c)$ theory \cite{ArPlSh2},
however, we think it reasonable to 
identify these points with those in the Coulomb branch
where the curve degenerates to genus zero curve as stated in subsection 2.2.
As we shall see in the following our interpretation naturally fits
into the M theory consideration.

\subsection{Fivebrane configuration in M theory}

Let us describe how the above type IIA brane configuration is embedded in 
M theory in terms of
a fivebrane whose worldvolume is
${\bf R^{1,3}} \times \Sigma$.
The curve $\Sigma$ is identified with the Seiberg-Witten
curve that determines the solutions to Coulomb branch 
of the field theory.
As usual, we write $s=(x^6+i x^{10})/R$, $ t=e^{-s}$
where $x^{10}$ is the eleventh coordinate of M theory which is compactified
on a circle of radius $R$. Then the curve $\Sigma$ which describes
the $N=2$ $Sp(2 N_c)$ gauge theory with $N_f$ flavors
is given by an equation in $(v, t)$ space \cite{LaLoLo}
\beq
\label{sp}
t^2-2 \left( v^2 B_{2 N_c} (v^2, u_{2 k}) +
\La_{N=2}^{2 N_c+2-N_f} i^{N_f} \prod_{i=1}^{N_f} m_i \right) t 
+ \La_{N=2}^{2(2 N_c+2-N_f)} \prod_{i=1}^{N_f} (v^2-{m_i}^2)=0,
\eeq
where  $B_{2 N_c} (v^2, u_{2 k})$ is a degree $2 N_c$ polynomial in $v$
which is even under $v \rightarrow -v$,
and the coefficients
depend on the moduli $u_{2 k}$ as well as the quark mass  $m_i$.


In M theory, the D6-branes are 
Kaluza-Klein monopoles described by a multi-Taub-NUT space \cite{To}. 
To construct the Seiberg-Witten curves 
we do not need
the full details of the metric of this multi-Taub-NUT space
but only a description of the space in terms of its
one complex structure.
Such a description has been constructed in \cite{GiRy} \cite{LaLoLo} and 
the result is 
\beq
\label{spd6}
t z=\Lambda_{N=2}^{2(2 N_c+2-N_f)} \prod_{i=1}^{N_f} (v^2-{m_i}^2)
\eeq
in ${\bf C}^3$ for $Sp(2 N_c)$
where one of the coordinates of ${\bf C}^3$ is taken
to be $t$ of the curve (\ref{spcurve}) for simplicity.
The Riemann surface $\Sigma$ is embedded 
as a curve in this surface and
given by
\beq
\label{spsigma}
t+z= 2 \left( v^2 B_{2 N_c} (v^2,u_{2 k}) 
+ \La_{N=2}^{2 N_c+2-N_f} i^{N_f} \prod_{i=1}^{N_f} m_i \right).
\eeq


The surface (\ref{spd6}) becomes singular
when some $m_i$ take the same value.
Physically this corresponds to the coincident D6-branes
in the $x^4,x^5$ directions but they can be separated 
in the $x^6$ direction \cite{HaWi}.
The separations of the D6-branes in the $x^6$ direction correspond
to the resolution of the singularities \cite{Wi}.

When all bare masses are turned off,
the surface (\ref{spd6}) 
develops singularities of type $A_{2 N_f-1}$ 
at the point $t=z=v=0$ \footnote[2]{I would like to thank K. Hori for
his pointing out an error in our original argument based on \cite{AhOhTa} }.
By succession of blowing ups, we obtain a smooth 
complex surface which 
isomorphically maps onto the singular surface (\ref{spd6}) 
except at the inverse image of the 
singular points. 
Over each singular point, there exist $2 N_f-1$ rational curves 
$\CP^1$'s called the exceptional curves on this smooth surface.
We denote the rational curves by $C_1, C_2, \cdots , C_{2 N_f-1}$.
This resolved surface is described as follows.
It is covered by $2N_f$ complex planes $U_1, U_2, \cdots , U_{2 N_f}$
with coordinates $(t_1 =t, z_1), (t_2, z_2) ,\cdots , (t_{2 N_f}, z_{2 N_f}=z)$ 
which are mapped
to the singular surface by
\beq
U_i \ni (t_i, z_i) \mapsto  \left\{ \begin{array}{l}
                t= t^i_i z^{i-1}_i\\
                z = t_i^{2N_f-i} z_i^{2N_f+1-i}\\
                v= t_i z_i
                        \end{array} \right.
\eeq
The planes $U_i$ are glued together by $z_i t_{i+1} =1$ 
and $t_i z_i = t_{i+1}z_{i+1}$. 
We define the exceptional curve $C_i$ by the locus of 
$t_i=0$ in $U_i$ and $z_{i+1}=0$ in $U_{i+1}$.
The spacetime reflection due to the orientifold can be extended to
the resolved surface by considering the action
$t_i \rightarrow (-1)^{i+1} t_i$, $z_i \rightarrow (-1)^i z_i$.
Under the reflection, $C_{2 n-1}$ parametrized by $z_{2n-1}$ 
is rotated while $C_{2n}$ is invariant.
The positions of the $D6$-branes may 
be interpreted as the $2 N_f$ intersection points 
of the exceptional curves.

\subsection{Higgs Branch}

Now we would like to study the Higgs branch 
when all the bare masses are turned off.
In M theory,  the Higgs branch appears when the fivebrane
intersects with the D6-branes.
This is possible only
when the image (\ref{spsigma}) in the $t$-$z$-$v$ space of the curve
passes through the singular point $t=z=v=0$.
Thus, we must have $v^2 B_{2 N_c}(v^2)=0$ at $v=0$ but
this condition is always satisfied due to the vanishing bare mass.
However the extra $v^2$ factor has its origin in the bending and 
connecting fivebrane at $v=0$
by the effect of the charge of the orientifold \cite{LaLoLo}.
Therefore we must have 
$B_{2 N_c}(v^2)$ in the factorized form
\beq
B_{2 N_c}(v^2)=v^{2 r} 
(v^{2(N_c-r)}+s_2 v^{2(N_c-r-2)}+\cdots+s_{2(N_c-r)})\,,
\label{spnbc}
\eeq
where $r>0$.

We will examine the curve by separating it into
the one near the singularity $t=z=v=0$, 
and the one away from it.
Away from the singular point
$t=z=v=0$, we can consider the curve
as embedded in the original $t$-$z$-$v$ space
because there is no distinction from the resolved surface in this
region.
Because $v$ never
vanishes in this region of the curve,
we can safely divide the coordinates $t$ and $z$ by some
power of $v$. If $2 r\leq N_f$ they can be divided by $v^{2 r}$,
and we see that this piece of the curve is equivalent with the
generic curve
for the $Sp(2(N_c-r))$ gauge theory with $(N_f-2 r)$ flavors
and thus has genus $(N_c-r)$.

Near $t=z=v=0$, we should consider the resolved $A_{2 N_f-1}$ surface.
Thus, we must describe the curve in the $2 N_f$ patches as described
above. 
It is useful to remark that the higher
order terms $v^{2(r+2)},v^{2(r+3)},\ldots$ are negligible near $v=0$
compared to $v^{2 (r+1)}$.
Thus we can replace the defining equation
$t+z=v^2 v^{2r} (s_{2(N_c-r)}+\cdots)=0$ by $t+z=v^2 v^{2 r}$ 
where we set $s_{2(N_c-r)}=1$ for notational simplicity.
On the $i$-th patch $U_i$, the equation of the curve $\Sigma$ becomes
\beq
t_i^i z_i^{i-1}+t_i^{2 N_f-i} z_i^{2 N_f+1-i}=t_i^r z_i^r (t_i z_i)^2 .
\eeq
Thus we have this equation factorize as
\beqa
&&
t_i v_i^{i-1} (1+v_i^{2 N_f-2 i} z_i^{2}
-v_i^{2 r-i} z_i (v_i)^2 )=0,
\qquad i=1,\ldots,2 r
\nonumber\\[0.2cm]
&&
v_i^{2 r} (t_i v_i^{i-2 r-1}
+v_i^{2 N_f-2 r-i} z_i-(v_i)^2 )=0,
\qquad i=2 r+1,\ldots,2 N_f-2 r
\nonumber\\[0.2cm]
&&
v_i^{2 N_f-i} \! z_i
(t_i^{2}v_i^{2i-2 N_f-2}+1 
-t_i v_i^{i-2 N_f+2 r-1} (v_i)^2) \!=\! 0,\;
i=2 N_f-2 r+1,\ldots,2 N_f, 
\label{spfac}
\eeqa
where we define $v_i=t_i z_i$.
However we can further factor out  $t_i, t_i^2 z_i, (t_i z_i)^2,
t_i z_i^2$ and $z_i$ for $i=2r +1, i=2r +2, 
2r+3 \leq i \leq 2 N_f-2 r-2, i=2 N_f-2 r-1$ and $i=2 N_f-2 r$
respectively from (\ref{spfac}).
These factors are interpreted as the fivebrane fixed at $v=0$,
which may have no contribution to the
dimension of the Higgs branch.

In addition to the above components,
the curve consists of several components which correspond to the D4-branes
as a consequence of the factorized form (\ref{spfac}).
One component, 
which we call $C$, is the zero of the
last factor of (\ref{spfac}). 
This extends to the one in the region away from 
$t=z=v=0$ which we have already considered. 
The other components are 
the rational curves $C_{1}, \ldots, C_{2 N_f-1}$
with multiplicities.
As is evident from the
factorized form (\ref{spfac}), the component $C_i$ has multiplicity $\ell_i$
where
$\ell_i=i$ for $i=1,\ldots,2 r$, $\ell_i=2 r$ for $i=2 r+1,\ldots,2 N_f-2 r-1$,
and $\ell_i={2 N_f-i}$ for $i=2 N_f-2 r,\ldots, 2N_f-1$.
Note that the component $C$ intersects 
with $C_{2r +2 }$ and $C_{2 N_f-2 r-2}$ and 
the unmoving $\CP^1$ components are aligned
from $C_{2r +1}$ to $C_{2 N_f-2 r-1}$.
We also note that 
the branes configuration in the type IIA theory
with the $s$-rule is verified by 
the above M theory fivebrane configuration.

To count the dimension of the Higgs branch, remember
that once the curve degenerates and $\CP^1$ components are generated, they can move
in the $x^7, x^8, x^9$ directions \cite{Wi}.
This motion together with the integration of the chiral two-forms on such
$\CP^1$'s parameterizes the Higgs branch of the four-dimensional theory.
As in the type IIA picture,
taking into account the orientifolding,
we expect that $\CP^1$'s move in pairs.
The $\CP^1$'s which can not become pairs are fixed.
Thus we find that the complex dimension of the $r$-th Higgs branch is 
\beq
2 \sum_{i=1}^{2 N_f-1} \left[ \frac{l_i}{2} \right] =
4 \left( 2 \sum_{i=1}^{r-1} i +r \right)
+2 r (2 N_f-1-4 r) = 2 r(2 N_f-2 r-1)
\eeq
in agreement with (\ref{spdH}).

For the case $N_f \geq N_c+2$,
there is a special point in the Higgs branch root 
in which the maximal massless dyons
appear \cite{ArPlSh2}.
This special point in the Higgs branch root is 
in the $r^*$-th Higgs branch root where
$r^*=N_f-N_c-2$ and takes the color Casimirs such that
\beq
B_{2 N_c}(u_{2 N_c},v^2)=v^{2 N_c} 
+ \frac{1}{4} v^{2 r^*} \La_{N=2}^{2 (N_c-r^*)}
=v^{2 r^*} \pare{v^{2(N_c-r^*)}+\frac{1}{4} \La_{N=2}^{2 (N_c-r^*)} }.
\eeq
Then at this point the equation (\ref{spsigma}) factorizes as
\beq
\frac{1}{t} \pare{t-2 v^{2(N_c+1)}} 
\pare{t-\frac{1}{2} \La_{N=2}^{2(N_c-r^*)} v^{2(r^*+1)}}=0.
\label{spb}
\eeq
Namely, the infinite curve $C$ factorizes into two rational curves
--- $C_L$ and
$C_R$ corresponding to $t=2 v^{2(N_c+1)}$ 
and $t=\frac{1}{2} \La_{N=2}^{2(N_c-r^*)} v^{2(N_f-N_c-1)}$
respectively.
We note that these curves are invariant under the 
orientifold projection $v \rightarrow -v, t \rightarrow t$.
It is also noted that the two curves $C_L$ and $C_R$ intersect
at $(2N_c+2 -N_f)$ points.

\section{$N=1$ moduli space of $Sp(2 N_c)$ theory from M theory}

In this section we deform the fivebrane configuration in M theory
for $N=2$ field theory to the one for $N=1$ field theory.

First we will study the configuration
of fivebrane in M theory which corresponds to
adding only a mass term to the adjoint chiral multiplet 
in the $N=2$ vector multiplet.
In the Type IIA picture, 
this corresponds to rotating the one of the NS 5-branes.

Next we will consider the configuration
of fivebrane in M theory which corresponds to
more general perturbation (\ref{sppert}).
In the Type IIA picture, 
this corresponds to adding plural NS 5-branes and 
changing the relative orientation of the NS 5-branes.

\subsection{Rotated configurations}

Here we study the adjoint mass perturbation.
Let us introduce
a complex coordinate
\beq
    w = x^8 + i x^9.
\eeq
Before breaking the $N=2$ supersymmetry, the fivebrane is located at $w=0$.
Now we rotate only the left NS 5-brane toward the $w$ direction.
From the behavior of
two asymptotic regions which correspond 
to the left and right NS 5-brane 
with  $v \rightarrow \infty$,
this rotation leads to the boundary conditions 
\beqa
\label{spbdy-cond}
& & w \rightarrow \mu v \;\;\; \mbox{as}\;\;\; v \rightarrow 
\infty, \;\;\; t \sim v^{2 N_c+2}  \nonumber \\
& & w \rightarrow 0 \;\;\; \mbox{as}\; \;\; v \rightarrow 
\infty, \;\;\; t \sim
\Lambda_{N=2}^{2(2 N_c+2-N_f)}v^{2 (N_f-N_c-1)}.
\eeqa
We can identify $\mu$ as the mass of the adjoint chiral multiplet
by using the $R$-symmetries.

Such rotation is only possible at points in the moduli space 
at which all $\beta$-cycles on the curve $\Sigma$ are degenerate.
This is possible at the special points in the Coulomb branch 
or in the $r^*$-th Higgs branch root.
The possible fivebrane configuration which corresponds to 
the points in the Coulomb branch is 
studied in \cite{AhOhTa}.

Here we consider the fivebrane configuration corresponding to 
the special point in the $r^*$-th Higgs branch root
which plays the important role in the arguments of
demonstrating the  $N=1$ Non Abelian duality \cite{ArPlSh2}.

The rotation of the curve at this point is straightforward.
The component $C$ at this point
factorizes into two pieces ---
$C_L$ described by $t=2 v^{2 (N_c+1)}$, $w=0$ and $C_R$ described by
$t=\frac{1}{2} \La_{N=2}^{2(2 N_c-N_f+2)} v^{2(N_f-N_c-1)}$, $w=0$.
The rotation can be done just by
replacing $w=0$ for $C_L$ by
$w=\mu v$.
The curve is explicitly given by
\beq
\widetilde{C}_L\,
\left\{
\begin{array}{l}
t=2 v^{2 (N_c+1)}\\
w=\mu v
\end{array}
\right.
\qquad
C_R\,
\left\{
\begin{array}{l}
t=\frac{1}{2} \La_{N=2}^{2(2 N_c-N_f+2)} v^{2(N_f-N_c-1)}\\
w=0 .
\end{array}
\right.
\label{sprotb}
\eeq
Note that the curves (\ref{sprotb}) are invariant under the 
orientifold reflection 
$v \rightarrow -v, t \rightarrow t, w \rightarrow -w$.

We can take the limit $\mu \rightarrow \infty$
which corresponds to the $N=1$ supersymmetric QCD 
in view of field theory.
Following \cite{AhOhTa},
we should rescale $t$ by a factor $\mu^{2 N_c+2}$ and introduce
a new variable
\beq
\tilde{t}= \mu^{2 (N_c+1)} t.
\eeq
Using the rescaled variable, the spacetime is described by
\beq
\tilde{t} z=\mu^{2(N_{c}+1)}
\Lambda_{N=2}^{4N_{c}+4-2N_{f}}\prod_{i=1}^{N_{f}} (v^{2}-m_{i}^{2}).
\eeq
This equation defines a smooth surface in the limit
$\mu \rightarrow\infty$ provided the constant $\La_{N=1}$ given by
\beq
\La_{N=1}^{2(3(N_c+1)- N_f)} =\mu^{2(N_c+1)}
\La_{N=2}^{2(2 N_c+2-N_f)} 
\eeq
which is kept finite.
This relations is the same as the renormalization 
matching condition of the corresponding field theory.

We can take the limit $\mu \rightarrow \infty$ of 
the curves of the remaining Higgs branch root (\ref{sprotb})
and the curves become
\beq
C_L\,
\left\{
\begin{array}{l}
\tilde{t}=2 w^{2 (N_c+1)}\\
v=0
\end{array}
\right.
\qquad
C_R\,
\left\{
\begin{array}{l}
\tilde{t}=\frac{1}{2} \La_{N=1}^{2(3( N_c+1)-N_f)} v^{2(N_f-N_c-1)}\\
w=0 .
\end{array}
\right.
\label{sprotbl}
\eeq
Note that in this limit the extra rational curves may appear and
the dimension of the Higgs branch is increased as 
in field theory \cite{HoOoOz}.

\subsection{Brane configuration of more general $N=1$ theory}

Let us  now consider the configuration
of fivebrane in M theory which corresponds to
more general perturbation of the form $\Delta W$ (\ref{sppert}).

In the type IIA picture, 
this corresponds to the configuration consists of $N_c$ ${\rm NS'}$ 5-branes
and their mirror pairs located right to a NS 5-brane
which does not have mirror pair \cite{ElGiKuRaSc}.
We will take the 
$N_c$ ${\rm NS'}$ 5-branes to stretch in the $(v,w)$ coordinates.
In field theory the eigenvalues of the vev of $\p$ are
labeled by the minima of the superpotential (\ref{Wp}) classically
and they label the separation
of the ${\rm NS'}$ branes in the $v$ direction.

Next we will construct the configuration of M theory
fivebrane which corresponds to the perturbation 
$\Delta W$ (\ref{sppert}).
The left NS 5-brane corresponds to the asymptotic
region  $v\to \infty, t \sim v^{2(N_c+1)}$, the right $N_c$ ${\rm NS'}$ 5-branes
correspond to the asymptotic
region $v\to \infty, t\sim \La_{N=2}^{2(2N_c+2-N_f)} v^{2(N_f-N_c-1)}$.
The boundary conditions that we will impose are 
\beq
\begin{array}{ccl}
w & \rightarrow & \sum_{k=1}^{N_c} \mu_{2 k} v^{2 k-1}~~~~~{\rm as}~
v \rightarrow \infty,~
t \sim \La_{N=2}^{2(2N_c+2-N_f)} v^{2(N_f-N_c-1)}\\
w & \rightarrow & 0~~~~~~~~~~~~~~~~~~~~~{\rm as}~
v \rightarrow \infty,~
t \sim v^{2(N_c+1)} . \\[0.2cm] 
\end{array}
\label{cond}
\eeq

Alternatively, if the $N_c$ ${\rm NS'}$ 5-branes were located at the left and 
the NS 5-brane at the right 
the boundary conditions would read 
\beq
\begin{array}{ccl}
w & \rightarrow & \sum_{k=1}^{N_c} \mu_{2 k} v^{2 k-1}~~~~~{\rm as}~
v \rightarrow \infty,~
t \sim v^{2(N_c+1)}  \\
w & \rightarrow & 0~~~~~~~~~~~~~~~~~~~~~{\rm as}~
v \rightarrow \infty,~
t \sim \La_{N=2}^{2(2N_c+2-N_f)} v^{2(N_f-N_c-1)} . \\[0.2cm] 
\end{array}
\label{cond2}
\eeq

The motion of the D4-brane
between the NS-${\rm NS'}$ 5-branes is the degree of freedom corresponding to
the adjoint scalar field $\Phi$.
However this motion is not possible because NS 5-branes and ${\rm NS'}$ 5-branes 
are not parallel.
This will give rise to a potential which gives to the mass 
to $\p$, say $M_\p$.
It is expected that 
\beq
\pa w(v) / \pa v \sim M_\p  ~~~~~ {\rm at} ~~~~  v=\bra \p \ket ,
\label{bc1}
\eeq
where $w(v)$ is the position of the one of ${\rm NS'}$ 5-branes.
For $\Delta W=\frac{1}{2} \mu \tr \p^2$,
it is clear that (\ref{bc1}) holds since $w= \mu v$.
Eq. (\ref{bc1}) is also valid for more general $\Delta W$ and
consistent with boundary condition (\ref{cond}) 
since $M_{\p}$ is given by $\Delta W''(\p)$.

In the fivebrane configuration  $SU(2)_{7,8,9}$
is broken to $U(1)_{8,9}$ if the parameter $\mu_{2 k}$ 
is assigned the $U(1)_{4,5}\times U(1)_{8,9}$
charge $(2-4k,2)$.
We list below the charges of the coordinates and parameters.
\beq
\begin{array}{cccl}
&U(1)_{4,5}&U(1)_{8,9}&\\
v&2&0&\\
w&0&2&\\
t&2(N_c+1)&0&\\
z&2(N_c+1)&0&\\
\mu_k&2(1-2k)&2\\
\Lambda_{N=2}&2&0&
\end{array}
\label{splist}
\eeq
We can identify $U(1)_{45} = U(1)_R$ and $U(1)_{89}=
U(1)_J$.

From the field theory results,
we expect to be able to construct the fivebrane configuration 
for only special values of $u_k$'s.
Let us consider a perturbation of the form 
$\sum_{k=1}^{N_c-l'+1} \mu_{2 k} \frac{1}{2 k} \tr(\Phi^{2 k})$. 
The point in the moduli space of vacua that remains
as a vacuum after this perturbation
is the singular locus of the $N=2$ Coulomb branch where $l$ or more mutually
local dyons become massless in field theory point of view. 
In the M-theory picture,  
it is possible to construct the corresponding
fivebrane only when the $(v,t)$ curve degenerates 
to a genus $g \leq 2 N_c-2 l'$ curve.
It can be seen by that 
the boundary conditions (\ref{cond}) or (\ref{cond2})  
mean that $w$ is a meromorphic function of the $(v,t)$
which has a pole of order $2 N_c-2 l'+1$ at one point.
Such a function exists only when the $(v,t)$ curve
is equivalent to a genus $g \leq 2 N_c-2 l'$ curve.

Following the paper of \cite{DeOz}, 
we will find the possible fivebrane configurations 
with the boundary conditions (\ref{cond}).
Here it is assumed that 
the equation defining the $N=2$ curve,
\beq
t^2-2 C(v^2,u_{2 k}) t +G(v^2,u_{2 k}) =0,
\label{spc}
\eeq
where $C=v^2 B_{2 N_c} + \La_{N=2}^{2 N_c+2-N_f} 
i^{N_f} \prod_{j=1}^{N_f} m_j$ and 
$G=\La_{N=2}^{2( 2 N_c+2-N_f)} \prod_{i=1}^{N_f} (v^2-m_i^2)$,
remains unchanged.
We also assume that $w$ will be a rational function of
$t$ and $v$.

Using (\ref{spc}), we can rewrite $w(v,t)$ in the form
\beq
\label{au2}
w(t,v) = \frac{a(v) t + b(v)}{c(v) t + d(v)}.
\eeq
Let us denote the two solutions of (\ref{spc}) by $t_{\pm}(v)$
and consider a point in the $N=2$ moduli space
of vacua where the $(x,t)$ curve degenerates to a genus $N_c-l$ curve $(l \geq 1)$.
Hence we have that
\beq
v^2 y^2 = C^2(v^2)-  G(v^2) \equiv S^2(v) T(v^2) ,
\eeq
where 
\beq
S = v \prod_{i=1}^{l}(v^2-p_i^2), \;\;\;\;\;\;
T = \prod_{j=1}^{2 N_c+1-2l} (v^2-q_j^2).
\label{sppr1}
\eeq
Note that 
there is no poles for a finite value of
$v$ since there are no other infinite NS 5-branes in the type IIA
picture as in the $SU(N_c)$ case.
Thus, according to \cite{DeOz},
we get 
\beq
\label{spwdef}
w  =  N + \frac{H}{S} (t-C),
\eeq
where $N$ and $H$ are some polynomials in $v$ and should not depend on $t$, or
\beq
\label{spwdef1}
w(t_{\pm}(v),v)  =  N(v) \pm H(v) \sqrt{T(v)}.
\eeq

In the case of $Sp(2 N_c)$ gauge group theory,
there is the orientifold and
(\ref{spwdef}) has to be invariant under the reflection
$w \rightarrow -w,v \rightarrow -v, t \rightarrow t$.
We see that $C,S$ and $T$ transform 
to $C,-S$ and $T$ under this reflection respectively.
Thus we require that under this reflection $H$ and $N/v$ are invariant,
in other words $H=H(v^2)$.

Next we must impose the boundary conditions on (\ref{spwdef}).
As $v\rightarrow \infty$ and $t=t_{-} (v) \sim v^{2(N_c+1)}$, we want that 
$w \rightarrow 0$. This completely fixes $N$ as
\beq
\label{spau9}
N(v) = [ H(v^2) \sqrt{T(v^2)} ]_{+} ,
\eeq
where $[f(v)]_{+}$ denotes 
the part of $f(v)$ with non-negative powers of $v$,
in a power series expansion around $v=\infty$.
This form of $N$ is indeed odd degree in $v$
due to the fact that 
$\sqrt{T}=v^{2 N_c+1-2 l} \prod_i (1-b_j^2 / v^2)^{\frac{1}{2}}$.

In the other asymptotic region, $v \rightarrow \infty$ and
$t = t_{+}(v) \sim v^{2(N_f-N_c-1)}$, we see $w$ behaves as
\beq
\label{spau99}
w = [2 H(v^2) \sqrt{T(v^2)}]_{+} + {\cal O}(v^{-1}).
\eeq
If we take the $N=1$ perturbation 
of the form $\sum_{k=1}^{N_c-l'+1} \mu_{2 k} \tr\Phi^{2 k} / 2 k$
and $H$ as the degree $2 s$ polynomial in $v$,
the equation (\ref{spau99}) 
implies $s=l-l'$.
As a consequence of this and $s\geq 0$,
we find $l \geq l'$.
This clearly shows the relation
between the genus of the degenerate Riemann surface, and the
minimal power needed in the superpotential. 
In particular $l=1$ is the lowest values of $l$.

We note that imposing the other boundary condition (\ref{cond2})
simply corresponds
to the choice $N(v)=-[H(v^2) \sqrt{T(v)}]_{+}$. 
Finally, we note that $w$ satisfies the following important equation
\beq
\label{spau12}
w^2 -2 N w + N^2 - T H^2 = 0 .
\eeq

\subsection{Comparison to field theory}

In the following we explicitly construct the brane configuration
of the superpotential perturbation $\Delta W$ (\ref{sppert})
of the $N=2$ theory and compare the results to the field theory analysis
in section \ref{field}.

We will start with the $Sp(2 N_c)$ pure Yang-Mills theory.
We already know that the most general deformation of the fivebrane is
\beq
\label{spwpure}
w = N(v) + H(v^2) \frac{t-C(v^2)}{v \prod_{i=1}^{l} (v^2-p_i^2)} ,
\eeq
where $H(v^2)$ and $N(v)$ are arbitrary polynomials of $v$.
Consider the deformation of the right ${\rm NS'}$ 5-branes. 
We must impose the boundary conditions (\ref{cond}). 
As shown in the previous subsection,
the second boundary condition implies that $N$ has to be given by (\ref{spau9}).
\beq
N=\left[ H \prod_{j=1}^{2 N_c+1-2 l}(v^2-q_j^2)^{1/2} \right]_{+} ,
\eeq
where $[f(v)]_{+}$ denotes the  part of $f(v)$ with non-negative
powers of $v$.
From the first boundary condition in (\ref{cond})
the relation between $H(v^2)$ and the
values of $\mu_{2 k}$ can be determined by expanding $w$
as given in (\ref{spau99}) in powers of $v$. Using that
$t=2 C(v^2) + {\cal O}(v^{-2(N_c+1)})$ 
we find
\beq
w = 2 H(v^2) \frac{v^2 B_{2 N_c}(v^2)+i^{N_f} \prod_{i=1}^{N_f} m_i}
{v \prod_{i=1}^{l}(v^2-p_i^2) }
 + {\cal O}(v^{-1}) =  \sum_{k=1}^{N_c} \mu_{2 k} v^{2 k-1}
+ {\cal O}(v^{-1}),
 \label{speq}
\eeq
which determines $H(v^2)$ in terms of $\mu_{2 k}$. 
Note that the term proportional to $\prod_{i=1}^{N_f} m_i$ is 
of order ${\cal O}(v^{-3})$ and will be ignored in (\ref{speq}).

Expand $H(v^2)$ as
\beq
\frac{2 H(v^2)}{\prod_{i=1}^l (v^2-p_i^2)}=
\sum_{i=1}^l \frac{\omega_i}{(v^2-p_i^2)} ,
\label{spH}
\eeq
then (\ref{speq}) agrees with 
the field theory result (\ref{spau30}).
Eq. (\ref{speq}) together with (\ref{sppr1})
determines the $N=1$ moduli space 
of vacua after the perturbation and the dyon vevs.
We see that the M theory fivebrane describes correctly the fact that
only the singular locus of the $N=2$ Coulomb
branch is not lifted and rederive the equations that 
determine the vevs of the massless dyons along the singular locus.


In the case of $Sp(2 N_c)$ with $N_f$ flavors,
we can compute the dyon vevs 
along the singular locus which is not at the Higgs branch root similarly to 
the above consideration and the results agree with the field theory results
if $N_f \leq N_c+2$ since $t= 2 C(v^2)+{\cal O}(v^{2(N_f-N_c-1)})$. 
However, this does not
contradict (\ref{spau30}), because that result assumes the form
of the curve (\ref{spcurve}) which is not valid for $N_f> N_c+2$.
We have not considered the case $N_f>N_c+2$ in detail.


We can compute 
the dyon vevs at the roots of the Higgs branches,
for the cases with $N_f-2r-2 \leq N_c-r$.
Let us consider a point at the $r$-th Higgs branch root where 
the curve take the form (\ref{spyt}) following \cite{DeOz}.
Imposing the boundary conditions (\ref{cond}),
we easily get 
\beq
w=\sum_{k=1}^{N_c-r} \mu_{2 k} v^{2 k-1} + {\cal O}(v^{-1})
 =   2 H(v^2) \frac{v B_{2 (N_c-r)}(v^2)}{\prod_{i=1}^l  (v^2-p_i^2)} 
  + {\cal O}(v^{-1}),
\label{spbbf}
\eeq
which determines $H(v^2)$. Eq. (\ref{spbbf}) agrees with 
the field theory result (\ref{spbb}).


At the $r^*$-th Higgs branch,
the curve is represented by the fivebrane (\ref{sprotb} )
and remains two rational curves after the perturbation (\ref{sppert}) as
\beq
\widetilde{C}_L\, \left\{ \begin{array}{l} t=2 v^{2 (N_c+1)} \\
w= 0 \end{array} \right.
\qquad
C_R\, \left\{ \begin{array}{l}
t=\frac{1}{2} \La_{N=2}^{2(2 N_c-N_f+2)} v^{2(N_f-N_c-1)}\\
w=\sum_{k=1}^{N_c} \mu_{2 k} v^{2 k-1} .
\end{array}
\right.
\label{sprotb1}
\eeq
Thus we can see from 
the brane picture that the $r^*$-th Higgs branch
root is  not lifted for arbitrary values of the parameters $\mu_{2 k}$.


Here we will give the geometrical interpretation of the dyon vevs 
according to \cite{DeOz}.
Note that the dyon vev $m_i \tilde{m}_i$ is equal up to a factor of $\sqrt{2}$ 
to the difference
between the two finite values of $w$ (\ref{spwdef1}) as we take $v^2=p_i^2$.
And the singular $N=2$ curve (\ref{sppr}), (\ref{sp})
has a double point at $v^2=p_i^2,t=C(p_i^2)$.
After the perturbation $\Delta W$ (\ref{sppert}) 
this double point splits into two
separate points in $(v,t,w)$ space, and the distance between the points
in the $w$ direction is exactly the vev
of the dyon that became massless at this point in the $N=2$
theory. This provides a simple geometrical interpretation of
the dyon vevs in the brane picture.


The meson vevs are also obtained as the 
values of $w$ at $t=0, v^2=m_i^2$ following \cite{HoOoOz} \cite{DeOz}.
Let us now compute the finite values of $w$ at $t=0, v^2=m_i^2$ and
compare to the meson vevs for the case with one massless dyon,
in other words $l=1$ and $H$ is a constant.
Therefore we have that
\beq
\label{spab3}
C(v^2)^2 - \Lambda_{N=2}^{2 (2 N_c+2- N_f)} \prod_{i=1}^{N_f} (v^2-m_i^2) = 
v^2 (v^2-p^2)^2 T(v^2).
\eeq
and the function $w$ is given by 
\beq
w = [H \sqrt{T(v^2)}]_{+} \pm H \sqrt{T(v^2)} .
\eeq
Assuming $N_f \leq N_c+2$, 
we simplify the $\sqrt{T(v^2)}$ as 
\beq
\sqrt{T(v^2)} = \frac{C(v^2)}{v(v^2-p^2)} + {\cal O}(v^{-1}).
\eeq
We can always decompose 
\beq
\label{spab4}
C(v^2) = C(0)+\frac{v^2}{p^2} \pare{C(p^2)-C(0)} +
v^2 (v^2-p^2) \tilde{C}(v^2) ,
\eeq
for some $2 N_c-2$ degree polynomial $\tilde{C}$, and we
see 
\beq
\label{spab2}
\sqrt{T(v^2)} = v \tilde{C}(v^2) + {\cal O}(v^{-1}) \quad \longrightarrow
\quad  [\sqrt{T(v^2)}] = v \tilde{C}(v^2).
\eeq
Using (\ref{spab3}) and (\ref{spab4}) 
we get
\beq
\sqrt{T(m_i^2)} =   \frac{p^2 C(0)+m_i^2 (C(p^2)-C(0))}{p^2 m_i(m_i^2-p^2)} 
+ m_i \tilde{C}(m_i^2).
\eeq
Thus we obtain the finite value of $w_{(i)}$ as
\beq
\label{spbb9}
w_{(i)} \equiv w(v \rightarrow m_i) =
\frac{H}{p^2} \pare{ \frac{C(0)}{m_i} -\frac{m_i}{m_i^2-p^2} C(p^2)}.
\eeq
It is noted that the reflection due to the orientifold which change the sign 
of $w$ and $m_i$ changes the sign of $w_{(i)}$.
We can evaluate $C(0)$ and $C(p^2)$ from (\ref{spab3}).
We can also compute $p$ and $H$ from the asymptotic behavior
of $w$ for large $v$ and $t \sim v^{2(N_c+1)}$ and the results are
$2H = \mu_{2 N_c}$, $p^2 =\frac{\mu_{2(N_c-1)}}{\mu_{2 N_c}} -s_2$.
Then we find 
\beq
w_{(i)} = \frac{1}{2} \mu_{2 N_c} \frac{1}{p^2} \Lambda_{N=2}^{2 N_c+2-N_f}
\pare{ 
\frac{i^{N_f} \prod_{j=1}^{N_f} m_j}{m_i} \pm \frac{m_i}{p^2-m_i^2}
\prod_{j=1}^{N_f} (p^2-m_j^2)^{\frac{1}{2}} }.
\label{spWi}
\eeq

Here we will assume that 
if we take the perturbation $\Delta W =\sum_{i=1}^{N_c} \mu_{2 k} \tilde{u}_{2 k}$
we should 
replace $\mu_{2k}$ for boundary conditions (\ref{cond}) by $\tilde{\mu}_{2k}$.
This assumption is natural in the sense of the arguments in subsection 4.2.
Therefore taking the superpotential (\ref{wsp}), 
we can easily see that $p^2=\frac{\mu_{2 (N_c-1)}}{\mu_{2 N_c}}$
and the dyon condensation derived from 
the fivebrane in M theory still agrees
with field theory result.
Then comparing (\ref{spWi}) and (\ref{spbb8}) we see that
up to a factor of $\sqrt{2}$,
the values of $w$ at $t=0,v^2=m_i^2$ are  exactly the eigenvalues of 
meson vevs derived from the field theory taking the appropriate 
basis of Casimirs and assuming that
the low energy effective superpotential (\ref{wsp1}) is exact.

We can compute the meson vev for the case with more than
one massless dyon.
Following \cite{DeOz},
we find that
\beqa
w_{(i)} & =& m_i \La_{N=2}^{2 N_c+2-N_f} 
\left(
\frac{H(0)}{\prod_k (-p_k^2)} 
\frac{\prod_{f=1}^{N_f} (-m_f^2)^{\frac{1}{2}}}{-m_i^2} \right. \CR
& & \hspace{3cm} 
\left. \pm \sum_{j=1}^{l} \frac{H(p_j^2)}{p_j^2 \prod_{k \neq j}  (p_j^2-p_k^2)} 
\frac{\prod_{f=1}^{N_f} (p_j^2-m_f^2)^{\frac{1}{2}}}{p_j^2-m_j^2} \right) \CR
& =& -\frac{\pa}{\pa m_i} \pare{ \La_{N=2}^{2 N_c+2-N_f} \sum_{j=1}^{l} 
\frac{\omega_j}{p_j^2} \pare{  \sq{\det(-m)}
\pm \sq{\det(p_j-m)}  }}.
\label{spWi1}
\eeqa

We can also obtain the fivebrane configuration at the point in the moduli space of 
vacua with maximal number of mutually local massless dyons which implies
$l=N_c$ as in \cite{DeOz}, but here we omit this discussion for the brevity.

\section{Theories with $SO(2 N_c)$ gauge group}

The procedure discussed above can be also applied to the other classical 
gauge groups.
In this section we study the $SO(2 N_c)$ gauge group case.

\subsection{Field theory analysis of $SO(2 N_c)$ gauge theory}

Let us consider $N=2$ supersymmetric gauge theory 
with the gauge group $SO(2 N_c)$ and $N_f$ quark hypermultiplets 
in the fundamental representation $Q_a^i$, $i=1, \ldots, 2 N_f$.
Here $a=1, \ldots,2 N_c$ is color indice.
The scalar chiral multiplet
$\p_{ab}$ is the $2N_c \times 2N_c$ antisymmetric tensor 
and the $N=2$ superpotential takes the form 
\beq
W=\sqrt{2} J_{ij} Q_a^i \p_{ab} Q_b^j +\sqrt{2} m_{ij} Q_a^i Q_a^j,
\label{soW}
\eeq
where $J={\rm diag}(i\sigma_2, \cdots, i\sigma_2)$. 
From the $N=2$ supersymmetry we can take
the quark mass matrix as
$m={\rm diag}(m_1 \sigma_1, \cdots, m_{N_f} \sigma_1)$
where $\sigma_1=\pmatrix{0 & 1 \cr 1 & 0}$.

The flavor symmetry is $Sp(2 N_f)$ 
in addition to $SU(2)_R \times U(1)_R$ R-symmetry group for massless quarks.
The instanton factor is proportional to $\La^{4 N_c-4-2 N_f}_{N=2}$ 
and the $U(1)_R$ symmetry is anomalous and is broken down 
to $\Z_{4 N_c-4-2 N_f}$.

The Coulomb branch is $N_c$ complex dimensional and 
is parametrized by the gauge invariant order parameters 
\beqa
u_{2 k} &=& \left\langle \frac{1}{2 k} {\rm Tr} \left( \p^{2k} \right) 
\right\rangle, \;\;\; k=1, \ldots ,N_c-1 \CR
P &=& \left\langle {\rm Pf \p} \right\rangle.
\label{sogi}
\eeqa
Hereafter  we use 
$u_{2 k}={\rm Tr} ( \p^{2 N_c} )=
{\rm Pf \p}^2+\ldots$ instead of ${\rm Pf \p}$
because it is difficult to interpret 
${\rm Pf \p}$ as the boundary conditions of the fivebrane in M theory.
The Coulomb branch parametrizes a family of 
genus $N_c$ hyperelliptic curves \cite{ArSh} \cite{Ha}
\beq
y^2=x \pare{ B_{2 N_c}(x,u_{2 k})^2 
-x^2 \La^{2(2 N_c-2)-2 N_f}_{N=2} \prod_{i=1}^{N_f} (x-m_i^2) },
\label{socurve}
\eeq
where $B_{2 N_c}$ is a degree $N_c$ polynomial in $x$ with coefficients 
that depend on the gauge invariant order parameters $u_{2 k}$ 
and $m_i$.
For $N_f < N_c-2$ the polynomial $B_{2 N_c}$ is given \cite{Ha} by
\beq
B_{2 N_c}(x)=\sum_{i=0}^{N_c} s_{2 i} x^{N_c-i} .
\eeq

Hereafter we will consider the case with 
the vanishing bare quark mass for simplicity.
There are two types of the gauge invariants which are constructed from 
$Q$, the meson fields $M^{ij}=Q^i J Q^j$ and the baryon fields
$B^{[i_1 \ldots i_{N_c}]} = 
Q_{a_1}^{i_1} \cdots Q_{a_{N_c}}^{i_{N_c}} \E_{a_1  \ldots a_{N_c}}$.
The baryon field is defined for $N_f \geq N_c$.
The Higgs branches are classified by an integer $r$ 
such that $0 \leq r \leq {\rm min} \{ N_c,\frac{N_f}{2} \} $ \cite{ArPlSh2}.
The $r$-th Higgs branch has complex dimension $2 r (2 N_f-2 r+1)$
and emanating from the  $N_c-r$ dimensional submanifold of the Coulomb branch.
The baryon field has non vanishing vev 
only when $r=N_c \leq  N_f / 2$.
The effective theory along the root of the $r$-th Higgs branch
is $SO(2 r) \times U(1)^{N_c-r}$ with $N_f$ massless quarks 
which are neutral with respect to any $U(1)$ factor.
There are special points along the root where such massless matters exist.
In particular in $r^*$-th Higgs branch where $r^*=N_f-N_c+2$,
the maximal singularity occurs at which the curve takes the form
\beq
y^2=x^{2 r+1} \left( x^{N_c-r}+\frac{1}{4} \La^{2(N_c-r)} \right)^2,
\label{somax}
\eeq
and $N_c-r^*$ hypermultiplets become simultaneously massless.
Note that in the $r^*$-th Higgs branch the baryon field always has 
vanishing vev.


We consider the perturbation $\Delta W$ (\ref{sppert}) to the 
$N=2$ superpotential (\ref{soW}) 
\beq
W=\sqrt{2} J Q \Phi Q +\sqrt{2} m Q Q+ \Delta W.
\label{soWp}
\eeq
If only the mass perturbation is present,
only the special points in the $r^*$-th Higgs branch root 
and the point in the Coulomb branch where the curve degenerate
to the genus zero curve also remains as vacua as in the case of $Sp(2 N_c)$.

Now we consider $N=2$ pure $SO(2 N_c)$ Yang-Mills theory
perturbed by the superpotential $\Delta W$ (\ref{sppert}).
Near the singularities of the moduli space of vacua,
the low energy superpotential is also given 
by (\ref{spWp})
which is exact from the holomorphy and global symmetry arguments.
The list of charges under $U(1)_R$ and $U(1)_J \subset SU(2)_R$ is
same as (\ref{list1}) in $Sp(2 N_c)$ case.
And the equation of motion (\ref{vac1}) and (\ref{vac}) also hold in this case.
Let us consider a point in the moduli space where 
$l$ mutually local dyons are massless.
At this point the curve takes the form 
\beq
y^2=x \pare{ {B_{2 N_c}(x)}^2  -x^2 \La_{N=2}^{4 (N_c-1)} }=
\prod_{i=1}^{l}(x-p_i^2)^2
\prod_{j=1}^{2 N_c-2l+1} (x-q_j^2)
\label{sopr}
\eeq
with $p_i$ and $q_j$ distinct.
The equation of motion implies that
there will be a 
complex $N_c-l$  dimensional moduli space of $N=1$ vacua which remains
after the perturbation assuming the matrix
$\partial a_i / \partial u_{2 k}$ is non-degenerate.

Computation of the matrix $\partial a_i/\partial u_{2 k}$
can be done as in $Sp(2 N_c)$ case and using this 
we find the relation between the parameters $\mu_{2 k}$ and 
the dyon vevs $m_i \tilde{m}_i$ as
\beqa
\sum_{k=1}^{N_c} \mu_{2 k} v^{2 k-1}  & = &
 v B_{2 N_c}(v^2) \sum_{i=1}^l  \frac{\omega_i}{(v^2-p_i^2)} 
  + {\cal O}(v^{-1}) \CR
&=& v B_{2 N_c}(v^2) \frac{2 H(v^2)}{\prod_{i=1}^l (v^2-p_i^2)}+ {\cal O}(v^{-1}),
\label{soau30}
\eeqa
where we identify $x$ as $v^2$.
Here we define $\omega_i$ and $H$ from (\ref{spwi}) and (\ref{spH}).

In $N=1$ pure Yang-Mills theory with the gauge group $SO(2 N_c)$,
there are $2 N_c-2$ massive vacua where the discrete $\Z_{4 N_c-4}$
$R$-symmetry is spontaneously broken to $\Z_2$.
This vacuum corresponds to the genus zero curve and generically is not lifted. 


The computations of the dyon vevs in 
the the case of the $SO(2 N_c)$ supersymmetric QCD are straightforward
and the results are similar to the $Sp(2 N_c)$ case.
The vev of the meson field $Q Q$ 
along the singular locus of the Coulomb branch
is generated by the non perturbative dynamics of 
the $N=1$ theory and become nonzero.
We will compute this vev below.

We take a tree-level superpotential as
\beq
W=\sum_{n=1}^{N_c-1} \mu_{2 n} u_{2 n} + \mu_{2 N_c} s_{2 N_c}
+\sqrt{2}  J Q \p Q
+\sqrt{2}  m Q Q.
\label{wso}
\eeq
Here we choose $s_{2 N_c}$ instead of $u_{2 N_c}$ as in $Sp(2 N_c)$ case.
Due to this choice, we can easily obtain 
the low energy effective superpotential as
\beq
W(\mu_{2 k},m_i)
 = \sum_{n=1}^{N_c-1} \mu_{2 n} u_{2 n}^{cl} 
+ \mu_{2 N_c} s_{2 N_c}^{cl} 
  \pm \mu_{2(N_c-1)} \La_{N=2}^{2 N_c-2- N_f} 
\prod_{i=1}^{N_f} (p_1^2-m_i^2)^{\frac{1}{2}}  ,
\label{wso1}
\eeq
where $p_1^2=\frac{\mu_{2 (N_c-1)}}{\mu_{2 N_c}}$.
Indeed the vev of $s_{2 k}$ calculated from (\ref{wso1}) is
\beq
\langle s_{ 2 k} \rangle  =  s_{ 2 k}^{cl} (g) 
\pm \D_{k,N_c-1} \frac{2 A(p_1) +p_1^2 A'(p_1)}{2 \sqrt{A(p_1)}}
\mp \D_{k,N_c}  \frac{p_1^4 A'(a_1)}{2 \sqrt{A(p_1)}} ,
\label{vso1}
\eeq
where  $A(v) \equiv \La^{4( N_c-1)-2 N_f} \prod_{i=1}^{N_f} (v^2-m_i^2)
 =A(-v)$ and
$A'(v)=\frac{\pa}{\pa v^2} A(v)$
and these vev reproduces the singularities 
in the moduli space of the $N=2$ curves correctly.
Therefore we get the vev of the meson to derivate (\ref{wso1}) by $m_i$ as
\beq
\bra M^{ii} \ket = \mp \frac{1}{\sqrt{2}} \mu_{2 N_c}  p_1^2  
\La_{N=2}^{2 N_c-2-2 N_f} 
m_i \frac{ \prod_{i=1}^{N_f} (p_1^2-m_i^2)^{\frac{1}{2}} }{p_1^2-m_i^2}  .
\label{sobb8}
\eeq

The calculation of the dyon condensation under the perturbation (\ref{wso})
is similar to the $Sp(2N_c)$ case and the only modification is 
replacing  $\mu_{2 k}$ by $\tilde{\mu}_{2 k}$.

\subsection{$N=2$ Higgs branch of $SO(2 N_c)$ theory from M theory}

We will consider the moduli space of vacua of 
$N=2$ supersymmetric QCD with the gauge group $SO(2 N_c)$ and
its deformation by the superpotential (\ref{sppert})
using the M theory fivebrane.

The brane configuration in type IIA in which the four dimensional $SO(2 N_c)$
gauge theory is realized is almost same as $Sp(2 N_c)$ case but 
the Chan-Paton wavefunction of the vector is taken to be antisymmetric.

The $r$-th Higgs branch 
corresponds to $2 (N_c - r)$  D4-branes suspended between
the two NS 5-branes and $2 r$ D4-branes
broken on the D6-branes.
The $r$-th Higgs branch shares $(N_c-r)$ complex
dimensions with the Coulomb branch,
corresponding to the gauge group $SO(2(N_c-r))$.
Following what we have discussed in section 3,
we obtain the complex dimension of the $r$-th Higgs
branch to be
\beq
2 \sum_{l=1}^r \left[ 2 N_f-(4 l-3)) \right] = 2r(2 N_f-2 r+1)
\label{sodH}
\eeq
in agreement with the field theory results.


We embed the type IIA brane configuration in 
M theory fivebrane configuration.
The curve $\Sigma$, describing
the $N=2$ $SO(2 N_c)$ gauge theory with $N_f$ flavors,
is given by an equation in $(v, t)$ space \cite{LaLoLo}
\beq
\label{so}
v^2 t^2-2 B_{2 N_c}(v^2, u_{2 k}) t+
\La_{N=2}^{2( 2 N_c-2- N_f)} v^2 \prod_{i=1}^{N_f} (v^2-{m_i}^2)=0,
\eeq
where  $B_{2 N_c}(v^2, u_{2 k})$ is a degree $2 N_c$ polynomial in $v$,
and has the form 
$v^{2 N_c} + \cdots$ with only even degree in $v$.


We include the D6-branes to the above M theory fivebrane configuration. 
This situation is described by the surface
\beq
\label{sod6}
t z=\Lambda_{N=2}^{2(2 N_c-2-N_f)} \prod_{i=1}^{N_f} (v^2-{m_i}^2)
\eeq
in ${\bf C}^3$ for $SO(2 N_c)$. 
The Riemann surface $\Sigma$ is embedded 
as a curve in this curved surface and
is given by
\beq
\label{sosigma}
v^2 (t+z) = 2 B_{2 N_c}(v^2, u_{2 k}).
\eeq
Note that the $v^2$ factor in (\ref{sosigma}) has its origin in the 
fivebrane which is bent and 
infinitely extended along the orientifold
by the effect of the charge of the orientifold \cite{LaLoLo}.
The resolution of the singularities of (\ref{sod6}) 
is needed when some $m_i$ are coincident.
We can carry out this resolution as in $Sp(2 N_c)$ case.

Next we will study the Higgs branch 
when all the bare masses are turned off.
The Higgs branch appears when the fivebrane
intersects with the D6-branes.
Thus we have $B_{2 N_c}(v^2)$ in the factorized form
\beq
B_{2 N_c}(v^2)=v^{2 r} 
(v^{2(N_c-r)}+s_2 v^{2(N_c-r-2)}+\cdots+s_{2(N_c-r)})\,,
\label{sonbc}
\eeq
where $r>0$.

Away from the singular point
$t=z=v=0$,
the curve is equivalent with the
generic curve for the $SO(2(N_c-r))$ gauge theory with $(N_f-2 r)$ flavors
and thus has genus $(N_c-r)$
if $2 r\leq N_f$.
From (\ref{sosigma}) and (\ref{sonbc}),
one of the components of the curve, $C_0$, lies on $v=0$.

Near $t=z=v=0$, we can replace the defining equation
$v^2(t+z)=v^{2 r} (s_{2(N_c-r)}+\cdots)=0$ by $v^2(t+z)=v^{2 r}$.
On the $i$-th patch $U_i$, the equation of the curve $\Sigma$ becomes
\beq
(t_i z_i)^2 (t_i^i z_i^{i-1}+t_i^{2 N_f-i} z_i^{2 N_f+1-i})
=t_i^r z_i^r .
\label{sode}
\eeq
Thus we have this equation factorize as
\beqa
&&
t_i v_i^{i-1} (v_i^2 (1+v_i^{2 N_f-2 i} z_i^{2})
-v_i^{2 r-i} z_i )=0,
\qquad i=1,\ldots,2 r
\nonumber\\[0.2cm]
&&
v_i^{2 r} (v_i^2( t_i v_i^{i-2 r-1}
+v_i^{2 N_f-2 r-i} z_i)-1 )=0,
\qquad i=2 r+1,\ldots,2 N_f-2 r
\nonumber\\[0.2cm]
&&
v_i^{2 N_f-i}z_i
(v_i^2( t_i^{2}v_i^{2i-2 N_f-2}+\!1\! )
-t_i v_i^{i-2 N_f+2 r-1} )=0,
\; i=2 N_f\!-\!2 r+1,\ldots,2 N_f,
\label{sofac}
\eeqa
where we define $v_i=t_i z_i$.
However 
we can further factor out 
$t_i^2 z_i^2, t_i z_i^2, z_i, t_i, t_i^2 z_i$ and $ t_i^2 z_i^2  $ 
for $i \leq 2r-2, i=2r-1, i=2r, i=2 N_f-2 r+1, i=2 N_f-2 r+2$ and 
$i \geq 2 N_f-2 r+3$
respectively from (\ref{sofac}).
The components represented by these factors together with $C_0$
are interpreted as the infinitely extending fivebrane along the orientifold.
These components of the curve may have no contribution to the
dimension of the Higgs branch.

Thus we find the curve consists of
$C$, which is the zero of the last factor of (\ref{sofac}) and
extends to infinity, 
and the rational curves $C_1, \ldots , C_{2 N_f-1}$
with multiplicities.
From the factorized form (\ref{sofac}), 
the component $C_i$ has multiplicity $\ell_i$ where
$\ell_i=i$ for $i=1,\ldots,2 r$, $\ell_i=2 r$ for $i=2 r+1,\ldots,2 N_f-2r-1$
and $\ell_i=2 N_f-i$ for $i=2 N_f-2 r,\ldots,2 N_f-1$.
Note that the component $C$ intersects 
with $C_{2r -2 }$ and $C_{2 N_f-2 r+2}$.

As in the IIA picture,
it is expected that 
an additional $\CP^1$ are appeared in $C_{k}$ where $k$ is an odd integer.
Taking into account the orientifolding,
$\CP^1$'s must move in pairs.
Therefore if we define $o_k$ as $o_k=0$ for $k$ even and $o_k=1$ for $k$ odd,
the complex dimension of the $r$-th Higgs branch is obtained as
\beq
4 \sum_{i=1}^{2 N_f-1} \left[ \frac{l_i+o_i}{2} \right]  =
8 \sum_{i=1}^{r} i +2 r (2 N_f-1-4 r) = 2 r(2 N_f-2 r+1)
\eeq
in agreement with (\ref{sodH}).
It is noted that 
the fivebrane configuration subjected to the $s$-rule in the type IIA theory
is verified by the above M theory fivebrane configuration.

For the case of $N_f \geq N_c-2$,
there is a special point in the Higgs branch root.
This special point in the Higgs branch root is 
in the $r^*$-th Higgs branch root where
$r^*=N_f-N_c+2$ and takes the color Casimirs such that
\beq
B_{2 N_c}(u_{2 N_c},v^2)=v^{2 N_c} 
+ \frac{1}{4} v^{2 r^*} \La_{N=2}^{2 (N_c-r^*)}
=v^{2 r^*} \pare{v^{2(N_c-r^*)}+\frac{1}{4} \La_{N=2}^{2 (N_c-r^*)} }.
\eeq
Then at this point the equation (\ref{sosigma}) factorizes as
\beq
\frac{v^2}{t} \pare{t-2 v^{2 N_c-2}} 
\pare{t-\frac{1}{2} \La_{N=2}^{2(N_c-r^*)} v^{2(r^*-1)}}=0.
\label{sob}
\eeq

\subsection{$N=1$ moduli space of $SO(2 N_c)$ theory from M theory}

In the same way as the case with $Sp(2 N_c)$ group,
we will consider the deformation of the fivebrane configuration in M theory
for the $N=2$ $SO(2 N_c)$ gauge field theory
to the one for the $N=1$ field theory.

First we study the adjoint mass perturbation.
The boundary conditions we impose are 
\beqa
\label{sobdy-cond}
& & w \rightarrow \mu v \;\;\; \mbox{as}\;\;\; v \rightarrow 
\infty, \;\;\; t \sim v^{2 N_c-2}  \nonumber \\
& & w \rightarrow 0 \;\;\; \mbox{as}\; \;\; v \rightarrow 
\infty, \;\;\; t \sim
\Lambda_{N=2}^{2(2 N_c-2-N_f)} v^{2 (N_f-N_c+1)}.
\eeqa
We can identify $\mu$ as the mass of the adjoint chiral multiplet
by using the $R$-symmetries.

These conditions are only satisfied at points in the moduli space 
at which all $\beta$-cycles on the curve are degenerate.
This is possible at the special points in the Coulomb branch 
or in the $r^*$-th Higgs branch root.
The possible fivebrane configuration which corresponds to 
the points in the Coulomb branch is 
studied in \cite{AhOhTa2}.
The rotated curve for the $r^*$-th Higgs branch root is explicitly given by
\beq
\widetilde{C}_L\,
\left\{
\begin{array}{l}
t=2 v^{2 (N_c-1)}\\
w=\mu v
\end{array}
\right.
\qquad
C_R\,
\left\{
\begin{array}{l}
t=\frac{1}{2} \La_{N=2}^{2(2 N_c-N_f-2)} v^{2(N_f-N_c+1)}\\
w=0 .
\end{array}
\right.
\label{sorotb}
\eeq

The $N=1$ supersymmetric QCD limit, $\mu \rightarrow \infty$,
can be taken by introducing a new variable $\tilde{t}= \mu^{2 (N_c-1)} t$.
The equation (\ref{sod6}) defines a smooth surface in the limit
$\mu \rightarrow\infty$ provided the constant $\La_{N=1}$ given by
\beq
\La_{N=1}^{3(2 N_c-2)-2 N_f} =\mu^{2(N_c-1)} \La_{N=2}^{2(2(N_c-1)-N_f)} 
\eeq
is kept finite and this relation agrees with the field theory one.
We also take the limit $\mu \rightarrow \infty$ of 
the curves of the remaining Higgs branch root (\ref{sorotb})
and the curves become
\beq
C_L\,
\left\{
\begin{array}{l}
\tilde{t}=2 w^{2 (N_c-1)}\\
v=0
\end{array}
\right.
\qquad
C_R\,
\left\{
\begin{array}{l}
\tilde{t}=\frac{1}{2} \La_{N=1}^{3( 2 N_c-2)-2 N_f} v^{2(N_f-N_c+1)}\\
w=0 .
\end{array}
\right.
\label{sorotbl}
\eeq


Next let us consider the configuration
of fivebrane in M theory which corresponds to
more general perturbation of the form $\Delta W$ (\ref{sppert})
in a similar way.
The type IIA brane configuration is
same as the $Sp(2 N_c)$ case.

We will construct the configuration of M theory
fivebrane corresponding to it.
The boundary conditions which  we will impose are 
\beq
\begin{array}{ccl}
w & \rightarrow & \sum_{k=1}^{N_c} \mu_{2 k} v^{2 k-1}~~~~~{\rm as}~
v \rightarrow \infty,~
t \sim \La_{N=2}^{2(2N_c-2-N_f)} v^{2(N_f-N_c+1)}\\
w & \rightarrow & 0~~~~~~~~~~~~~~~~~~~~~{\rm as}~
v \rightarrow \infty,~
t \sim v^{2(N_c-1)} . \\[0.2cm] 
\end{array}
\label{socond}
\eeq

Under the assumption that the equation defining the $N=2$ curve 
remains unchanged and that $w$ will be a rational function of
$t$ and $v$,
we can proceed in the $SO(2 N_c)$ case along the consideration of \cite{DeOz}.
At the point in the moduli space where $l$ mutually local dyons are massless,
we can factorize the equation $y^2/v^6 =C^2-G=S^2 T$ where
$C= B_{2 N_c} /v^2$,
$G = \La_{N=2}^{2(2(N_c-1)-N_f)} \prod_{i=1}^{N_f} (v^2-m_i^2)$ and
\beq
S = \frac{1}{v^2} \prod_{i=1}^{l}(v^2-p_i^2) ,~~~
T = \prod_{j=1}^{2 N_c-2l} (v^2-q_j^2).
\label{sopr1}
\eeq
Then we find 
\beq
\label{sowdef}
w  =  N + \frac{H'}{S} (t-C),
\eeq
where $N$ and $H'$ are some polynomials in $v$.
Note that there are no poles for a finite value of $v$.
Since $C,S$ and $T$ are invariant under the reflection
$w \rightarrow -w,v \rightarrow -v, t \rightarrow t$,
we require that $H \equiv H'/v =H(v^2)$ and $N$ transforms to $-N$
under this reflection.

Now we must impose the boundary conditions (\ref{socond}) on 
the fivebrane configuration.
As $v\rightarrow \infty$ and $t=t_{-} (v) \sim v^{2(N_c-1)}$, we require that 
$w \rightarrow 0$. This determines $N$ as
\beq
\label{soau9}
N(v) = [ v H(v^2) \sqrt{T(v^2)} ]_{+} .
\eeq
In the other asymptotic region, $v \rightarrow \infty$ and
$t = t_{+}(v) \sim v^{2(N_f-N_c+1)}$, $w$ behaves as
\beq
\label{soau99}
w = [2 v H(v^2) \sqrt{T(v^2)}]_{+} + {\cal O}(v^{-1}).
\eeq
We note that $w$ satisfies the following important equation
\beq
\label{soau12}
w^2 -2 N w + N^2 - T v^2 H^2 = 0 .
\eeq


Let us focus on the $SO(2 N_c)$ pure Yang-Mills theory.
Using that $T=(t-C)/S$
and $t=2 C(v^2) + {\cal O}(v^{-2(N_c-1)})$,
the first boundary condition in (\ref{socond}) reads
\beq
w = 2 v H(v^2) \frac{B_{2 N_c}(v^2)}
{\prod_{i=1}^{l}(v^2-p_i^2) }
 + {\cal O}(v^{-1}) =  \sum_{k=1}^{N_c} \mu_{2 k} v^{2 k-1}
+ {\cal O}(v^{-1}),
 \label{soeq}
\eeq
which determines $H(v^2)$ in terms of $\mu_{2 k}$. 
The equation (\ref{soeq}) agree with 
the field theory equation (\ref{soau30}).


In the case of $SO(2 N_c)$ with $N_f$ flavors,
we can also compute the dyon vevs 
along the singular locus as in the previous calculations.
It is seen that at the $r^*$-th Higgs branch root
the curves (\ref{sorotb})
remain two rational curves after the rotation as
\beq
\widetilde{C}_L\, \left\{ \begin{array}{l} t=2 v^{2 (N_c-1)} \\
w= 0 \end{array} \right.
\qquad
C_R\, \left\{ \begin{array}{l}
t=\frac{1}{2} \La_{N=2}^{2(2 N_c-N_f-2)} v^{2(N_f-N_c+1)}\\
w=\sum_{k=1}^{N_c} \mu_{2 k} v^{2 k-1} .
\end{array}
\right.
\label{sorotb1}
\eeq
Thus we see from 
the brane picture that the $r^*$-th Higgs branch
root is not lifted for arbitrary values of the parameters $\mu_{2 k}$.


The meson vevs can be read from the 
values of $w$ at $t=0, v^2=m_i^2$ \cite{HoOoOz}.
We can compute the finite values of $w$ at $t=0, v^2=m_i^2$ and
compare these to the meson vevs for the case with one massless dyon,
in other words $l=1$ and $H$ is a constant.
Assuming $N_f<N_c-1$, 
we find
\beq
w_{(i)} = \pm \frac{1}{2} \mu_{2 N_c} \La_{N=2}^{2 N_c-2-N_f} p^2
m_i \frac{\prod_{j=1}^{N_f} (p^2-m_i^2)^{\frac{1}{2}}}{m_i^2-p^2}, ~~~
p^2=\frac{\mu_{2 (N_c-1)}}{\mu_{2 N_c}},
\label{soWi}
\eeq
which is agree with (\ref{sobb8}) up to a factor $\sqrt{2}$ .
Here we change the $\Delta W$ as (\ref{wso}) and the boundary conditions
as in the case of $Sp(2 N_c)$.
We also compute the meson vev for the case with more than
one massless dyon.
Following the previous calculations,
we can easily obtain that
\beqa
w_{(i)} &= & \pm \La_{N=2}^{2 N_c-2-N_f} m_i 
\sum_{j=1}^{l} \frac{p_j^2 H(p_j^2)}{\prod_{k \neq j} (p_j^2-p_k^2)} 
\frac{\prod_{f=1}^{N_f} (p_j^2-m_f^2)^{\frac{1}{2}}}{p_j^2-m_j^2} \CR
& =& \mp \frac{\pa}{\pa m_i} \pare{ \La_{N=2}^{2 N_c-2-N_f} \sum_{j=1}^{l} 
\sq{\det(p_j-m)} w_j p_j^2 }.
\label{soWi1}
\eeqa

\section{Theories with $SO(2 N_c+1)$ gauge group}

Let us now turn to the case of $SO(2 N_c+1)$ gauge group.
We will proceed in parallel with the case of $SO(2 N_c)$.

\subsection{Field theory analysis of $SO(2 N_c+1)$ gauge theory}

Consider $N=2$ supersymmetric gauge theory 
with the gauge group $SO(2 N_c+1)$ and $N_f$ quark hypermultiplets 
in the fundamental representation $Q_a^i$, $i=1, \ldots, 2 N_f$.
The $N=2$ superpotential takes the form (\ref{soW}).
The instanton factor is proportional to $\La^{2(2 N_c-1- N_f)}_{N=2}$ 
and the $U(1)_R$ symmetry is anomalous and is broken down 
to $\Z_{2(2 N_c-1- N_f)}$.

The Coulomb branch is $N_c$ complex dimensional and 
is parametrized by the gauge invariant order parameters 
\beq
u_{2 k} = \left\langle \frac{1}{2 k} {\rm Tr} \left( \p^{2k} \right) 
\right\rangle, \;\;\; k=1, \ldots ,N_c .
\label{soogi}
\eeq
The Coulomb branch parametrize a family of 
genus $N_c$ hyperelliptic curves \cite{ArSh} \cite{Ha}
\beq
y^2=x B_{2 N_c}(x,u_{2 k}) ^2
-x^2 \La^{2(2 N_c-1)-2 N_f}_{N=2} \prod_{i=1}^{N_f} (x-m_i^2),
\label{soocurve}
\eeq
where $B_{2 N_c}$ is a degree $N_c$ polynomial in $x$ with coefficients 
that depend on the gauge invariant order parameters $u_{2 k}$ 
and $m_i$.
For $N_f < N_c-1$ the polynomial $B_{2 N_c}$ is given by \cite{Ha}
\beq
B_{2 N_c}(x)=\sum_{i=0}^{N_c} s_{2 i} x^{N_c-i} .
\eeq

There are two types of the gauge invariants which are constructed from 
$Q$, the meson fields and the baryon fields, however,
the baryon field is defined for $N_f \geq N_c+1$.
The Higgs branches are classified by an integer $r$ 
such that $0 \leq r \leq {\rm min} \{ N_c,\frac{N_f-1}{2} \} $ \cite{ArPlSh2}.
The $r$-th Higgs branch has complex dimension $2 (2 r+1)( N_f- r)$
and emanating from the  $N_c-r$ dimensional submanifold of the Coulomb branch.
Classically the baryon field has non vanishing vev 
only when $r=N_c \leq \frac{ N_f-1}{2}$.
The effective theory along the root of the $r$-th Higgs branch
is $SO(2 r+1) \times U(1)^{N_c-r}$ with $N_f$ massless quarks 
which are neutral with respect to any $U(1)$ factor.
There are special points along the root where additional massless matters exist.
In particular in $r^*$-th Higgs branch where $r^*=N_f-N_c+1$,
the maximal singularity occurs at which the curve takes the form
\beq
y^2=x^{2 r^*+1} \left( x^{N_c-r^*}+\frac{1}{4} \La^{2(N_c-r^*)} \right)^2,
\label{soomax}
\eeq
and $N_c-r^*$ hypermultiplets become simultaneously massless.


We now consider the perturbation $\Delta W$ (\ref{sppert}) to the 
$N=2$ superpotential (\ref{soW}) with the gauge group $SO(2 N_c+1)$.
The effect of the mass perturbation is same as $SO(2 N_c)$ case.

First we study $N=2$ pure $SO(2 N_c+1)$ Yang-Mills theory
perturbed by the superpotential $\Delta W$ (\ref{sppert}).
Let us consider a point in the moduli space where 
$l$ mutually local dyons are massless where the curve takes the form 
\beq
y^2=x \pare{ {B_{2 N_c}(x)}^2  -x \La_{N=2}^{2 (2 N_c-1)} }=
\prod_{i=1}^{l}(x-p_i^2)^2
\prod_{j=1}^{2 N_c-2l+1} (x-q_j^2)
\label{soopr}
\eeq
with $p_i$ and $q_j$ distinct.
The equation of motion implies that
there will be a 
complex $N_c-l$  dimensional moduli space of the $N=1$ vacua which remains
after the perturbation assuming the matrix
$\partial a_i / \partial u_{2 k}$ is non-degenerate.

We can find the relation between the parameters $\mu_{2 k}$ and 
the dyon vevs $m_i \tilde{m}_i$ as
\beqa
\sum_{k=1}^{N_c} \mu_{2 k} v^{2 k-1}  & = &
 v B_{2 N_c}(v^2) \sum_{i=1}^l  \frac{\omega_i}{(v^2-p_i^2)} 
  + {\cal O}(v^{-1}) \CR
&=& v B_{2 N_c}(v^2) \frac{2 H(v^2)}{\prod_{i=1}^l (v^2-p_i^2)}+ {\cal O}(v^{-1}),
\label{sooau30}
\eeqa
where we identify $x$ as $v^2$.
Here we define $\omega_i$ and $H$ from (\ref{spH}) and (\ref{spwi}).
It is seen also in this case that
the vacua correspond to the curve (\ref{soocurve})
degenerates to a genus zero curve and generically is not lifted.
The computations of the dyon vevs in 
the the case of the $SO(2 N_c+1)$ supersymmetric QCD are straightforward
and the results are similar to the $Sp(2 N_c)$ case.

The vev of the meson field $Q Q$ 
along the singular locus of the Coulomb branch
is generated by the non perturbative dynamics of 
the $N=1$ theory and becomes nonzero.
We take a tree-level superpotential as
\beq
W=\sum_{n=1}^{N_c-1} \mu_{2 n} u_{2 n} + \mu_{2 N_c} s_{2 N_c}
+\sqrt{2}  J Q \p Q
+\sqrt{2}  m Q Q.
\label{wsoo}
\eeq
Here we choose $s_{2 N_c}$ instead of $u_{2 N_c}$ as in the $Sp(2 N_c)$ case.
Due to this, we can easily obtain 
the low energy effective superpotential as \cite{KiTeYa}
\beq
W(\mu_{2 k},m_i)
 = \sum_{n=1}^{N_c-1} \mu_{2 n} u_{2 n}^{cl} 
+ \mu_{2 N_c} s_{2 N_c}^{cl} 
  \pm \mu_{2 N_c} p_1 \La_{N=2}^{2 N_c-1- N_f} 
\prod_{i=1}^{N_f} (p_1^2-m_i^2)^{\frac{1}{2}}  ,
\label{wsoo1}
\eeq
where $p_1^2=\frac{\mu_{2 (N_c-1)}}{\mu_{2 N_c}}$.
Therefore we find the vev of the meson to derivate (\ref{wsoo1}) by $m_i$ as
\beq
\bra M^{ii} \ket = \mp \frac{1}{\sqrt{2}} \mu_{2 N_c}  p_1  
\La_{N=2}^{2 N_c-1-2 N_f} 
m_i \frac{ \prod_{i=1}^{N_f} (p_1^2-m_i^2)^{\frac{1}{2}} }{p_1^2-m_i^2}  .
\label{soobb8}
\eeq

\subsection{$N=2$ Higgs branch of $SO(2 N_c+1)$ theory from M theory}

We will consider the moduli space of vacua of 
$N=2$ supersymmetric QCD with the gauge group $SO(2 N_c+1)$ and
its deformations by the superpotential (\ref{sppert})
by analyzing M theory fivebrane.

The brane configuration in type IIA in which four dimensional $SO(2 N_c+1)$
gauge theory is realized is almost same as $SO(2 N_c)$ case.
Only difference between two cases is that 
there is an extra D4-brane on the orientifold four plane 
in the $SO(2 N_c+1)$ case.
This D4-brane can not move since it does not have a mirror partner.

The $r$-th Higgs branch 
corresponds to $2 (N_c - r)$  D4-branes suspended between
the two NS 5-branes and $2 r+1$ D4-branes
broken on the D6-branes.
The $r$-th Higgs branch shares $(N_c-r)$ complex
dimensions with the Coulomb branch,
corresponding to the gauge group $SO(2(N_c-r))$.
On the basis of our arguments in section 3, we get
the complex dimension of the $r$-th Higgs
branch
\beq
2 \sum_{l=1}^r \left[ 2 N_f-(4 l-3)) \right] +2(N_f-2 r) = 2(2r+1)(N_f-r)
\label{soodH}
\eeq
in agreement with the field theory results.
The last term on the lhs of (\ref{soodH}) comes from  
the extra D4-brane.


The curve $\Sigma$ in M theory, describing
$N=2$  $SO(2 N_c+1)$ gauge theory with $N_f$ flavors,
is given by an equation in $(v, t)$ space \cite{LaLoLo}
\beq
\label{soo}
v^2 t^2-2 v B_{2 N_c}(v^2, u_{2 k}) t+
\La_{N=2}^{2( 2 N_c-1- N_f)} v^2 \prod_{i=1}^{N_f} (v^2-{m_i}^2)=0,
\eeq
where  $B_{2 N_c}(v^2, u_{2 k})$ is a degree $2 N_c$ polynomial in $v$,
and has the form 
$v^{2 N_c} + \cdots$ with only even degree in $v$.


Including the D6-branes to the above M theory fivebrane configuration
is described by the surface
\beq
\label{sood6}
t z=\Lambda_{N=2}^{2(2 N_c-1-N_f)} \prod_{i=1}^{N_f} (v^2-{m_i}^2)
\eeq
in ${\bf C}^3$ for $SO(2 N_c+1)$. 
The Riemann surface $\Sigma$ is embedded in the surface and
given by
\beq
\label{soosigma}
v^2 (t+z) = 2 v B_{2 N_c}(v^2, u_{2 k}).
\eeq
Note that the factor $v^2$ on the lhs of (\ref{soosigma}) 
has its origin in the 
fivebrane which is bent and 
infinitely extended along the orientifold
by the effect of the charge of the orientifold and 
the factor of $v$ on the rhs of (\ref{soosigma}) has its origin in
the D4-brane without a mirror partner \cite{LaLoLo}.
The resolution of the singularities of (\ref{sood6}) 
is needed when some $m_i$ are coincident.
We can done this resolution as in $Sp(2 N_c)$ case.
However the spacetime reflection due to the orientifold 
is different from $Sp(2 N_c)$ case and 
it may be extended to
the resolved surface by considering the action
$t_i \rightarrow (-1)^{i} t_i$, $z_i \rightarrow (-1)^{i+1} z_i$.
In this case, $C_{2 n}$ is rotated by reflection due to the orientifold.

We now consider the Higgs branch where 
$B_{2 N_c}(v^2)$ factorizes as
\beq
B_{2 N_c}(v^2)=v^{2 r} 
(v^{2(N_c-r)}+s_2 v^{2(N_c-r-2)}+\cdots+s_{2(N_c-r)})\,.
\label{soonbc}
\eeq
Away from the singular point
$t=z=v=0$,
we see that the curve is equivalent with the
generic curve for the $SO(2(N_c-r))$ gauge theory with $(N_f-2 r-1)$ flavors
and thus has genus $(N_c-r)$
if $2 r\leq N_f$.
Note that the one of the components of the curve, $C_0$, lies on $v=0$.

Near $t=z=v=0$, we can replace the defining equation
$v^2(t+z)=v^{2 r+1} (s_{2(N_c-r)}+\cdots)=0$ by $v^2(t+z)=v^{2 r+1}$.
On the $i$-th patch $U_i$, the equation of the curve $\Sigma$ becomes
\beq
(t_i z_i)^2 (t_i^i z_i^{i-1}+t_i^{2 N_f-i} z_i^{2 N_f+1-i})
=t_i^r z_i^r (t_i z_i).
\label{soode}
\eeq
Thus we have this equation factorize as
\beqa
t_i v_i^{i-1} \left( v_i^2 (1+v_i^{2 N_f-2 i} z_i^{2})
-v_i^{2 r-i} z_i (v_i) \right)=0,
&&i=1,\ldots,2 r
\nonumber\\[0.2cm]
v_i^{2 r} \left(v_i^2( t_i v_i^{i-2 r-1}
+v_i^{2 N_f-2 r-i} z_i)-(v_i) \right)=0,
&&i=2 r+1,\ldots,2 N_f-2 r
\nonumber\\[0.2cm]
v_i^{2 N_f-i} \!z_i
\left(v_i^2( t_i^{2}v_i^{2i-2 N_f-2}+\!1 )
\!-\!t_i v_i^{i-2 N_f+2 r-1} (v_i) \right)=0,
&&\!\!\!\!\!\!i=2 N_f\!-\!2 r\!+\!1,\ldots,2 N_f,
\label{soofac}
\eeqa
where we define $v_i=t_i z_i$.
However 
we can further factor out 
$t_i^2 z_i^2$, $t_i z_i^2$, $t_i z_i, t_i^2 z_i$ and $ t_i^2 z_i^2  $ 
for $i \leq 2r-1$, $i=2r$, $2r+1 \leq i \leq 2N_f-2r$, $i=2 N_f-2 r+1$ and 
$i \geq 2 N_f-2 r+2$
respectively from (\ref{soofac}).
Among these components, the middle components $C_i$ $(2r+1 \leq i \leq 2N_f-2r-1)$ 
correspond to the extra D4-brane.
The other components represented by these factors together with $C_0$
are interpreted as the infinitely extending fivebrane along the orientifold
which may have no contribution to the
dimension of the Higgs branch.

Therefore we obtain the curve consists of
$C$, which extends to infinity, 
and the rational curves $C_1, \ldots, C_{2 N_f-1}$
with multiplicities.
From the factorized form, 
the component $C_i$ has multiplicity $\ell_i$ where
$\ell_i=i$ for $i=1,\ldots,2 r$, $\ell_i=2 r+1$ for $i=2 r+1,\ldots,2 N_f-2r-1$
and $\ell_i=2 N_f-i$ for $i=2 N_f-2 r,\ldots,2 N_f-1$.
Here the $\CP^1$'s coming from the extra D4-brane are 
included in the multiplicity $\ell_i$.
Note that the component $C$ intersects 
with $C_{2r -1 }$ and $C_{2 N_f-2 r+1}$.
From this we calculate the complex dimension of the $r$-th Higgs branch 
\beq
4 \sum_{i=1}^{2 N_f-1} \left[ \frac{l_i+o_i}{2} \right]   
=2(2 r+1)(N_f- r),
\eeq
which agrees with (\ref{soodH}).

For the case of $N_f \geq N_c-1$,
there is the special point in the $r^*$-th Higgs branch root where
$r^*=N_f-N_c+1$ and the color Casimirs are given by
\beq
B_{2 N_c}(u_{2 N_c},v^2)=v^{2 N_c} 
+ \frac{1}{4} v^{2 r^*} \La_{N=2}^{2 (N_c-r^*)}
=v^{2 ^*r} \pare{v^{2(N_c-r^*)}+\frac{1}{4} \La_{N=2}^{2 (N_c-r^*)} }.
\eeq
Then at this point the equation (\ref{soosigma}) factorizes as
\beq
\frac{v^2}{t} \pare{t-2 v^{2 N_c-1}} 
\pare{t-\frac{1}{2} \La_{N=2}^{2(N_c-r^*)} v^{2 r^*-1}}=0.
\label{soob}
\eeq

\subsection{$N=1$ moduli space of $SO(2 N_c+1)$ theory from M theory}

As in the case of $SO(2 N_c)$,
we consider the deformation of the fivebrane configuration in M theory
for the $N=2$ $SO(2 N_c+1)$ gauge field theory
to the one for the $N=1$ theory.

Let us start with the mass perturbation $\mu \tr{\p^2}/2$.
The boundary conditions we impose are 
\beqa
\label{soobdy-cond}
& & w \rightarrow \mu v \;\;\; \mbox{as}\;\;\; v \rightarrow 
\infty, \;\;\; t \sim v^{2 N_c-1}  \nonumber \\
& & w \rightarrow 0 \;\;\; \mbox{as}\; \;\; v \rightarrow 
\infty, \;\;\; t \sim
\La_{N=2}^{2(2 N_c-1-N_f)} v^{2 (N_f-N_c)+1}.
\eeqa
The conditions (\ref{soobdy-cond}) are only satisfied at points in the moduli space 
where the curve becomes completely degenerate.
This is possible at the special points in the Coulomb branch 
or in the $r^*$-th Higgs branch.
The possible fivebrane configuration which corresponds to 
the points in the Coulomb branch is 
studied in \cite{AhOhTa2}.
The rotated curves for the $r^*$-th Higgs branch are explicitly given by
\beq
\widetilde{C}_L\,
\left\{
\begin{array}{l}
t=2 v^{2 N_c-1}\\
w=\mu v
\end{array}
\right.
\qquad
C_R\,
\left\{
\begin{array}{l}
t=\frac{1}{2} \La_{N=2}^{2(2 N_c-N_f-1)} v^{2(N_f-N_c)+1}\\
w=0 .
\end{array}
\right.
\label{soorotb}
\eeq

As in the $SO(2 N_c)$ case the $N=1$ supersymmetric QCD limit can be taken 
if we define $\La_{N=1}$ as
\beq
\La_{N=1}^{3(2 N_c-1)-2 N_f} =\mu^{2 N_c-1} \La_{N=2}^{2(2 N_c-1- N_f)} .
\eeq
This relation agrees with the field theory one.
We also take the limit $\mu \rightarrow \infty$ of 
the curves of the remaining Higgs branch root (\ref{soorotb})
and the curves become
\beq
C_L\,
\left\{
\begin{array}{l}
\tilde{t}=2 w^{2 N_c-1}\\
v=0
\end{array}
\right.
\qquad
C_R\,
\left\{
\begin{array}{l}
\tilde{t}=\frac{1}{2} \La_{N=1}^{3( 2 N_c-1)-2 N_f} v^{2(N_f-N_c)+1}\\
w=0 .
\end{array}
\right.
\label{soorotbl}
\eeq


We will construct the configuration of M theory
fivebrane corresponding to more general perturbation 
of the form $\Delta W$ (\ref{sppert}).
The boundary conditions which we will impose are 
\beq
\begin{array}{ccl}
w & \rightarrow & \sum_{k=1}^{N_c} \mu_{2 k} v^{2 k-1}~~~~~{\rm as}~
v \rightarrow \infty,~
t \sim \La_{N=2}^{2(2N_c-1-N_f)} v^{2(N_f-N_c)+1}\\
w & \rightarrow & 0~~~~~~~~~~~~~~~~~~~~~{\rm as}~
v \rightarrow \infty,~
t \sim v^{2 N_c-1} . \\[0.2cm] 
\end{array}
\label{soocond}
\eeq
At the point in the moduli space where $l$ mutually local dyons are massless,
we can factorize the equation $y^2 /v^4=C^2-G=S^2 T$ where
$C = \frac{B_{2 N_c}}{v}$, 
$G = \La_{N=2}^{2(2 N_c-1-N_f)} \prod_{i=1}^{N_f} (v^2-m_i^2)$ and
\beq
S = \frac{1}{v} \prod_{i=1}^{l}(v^2-p_i^2) ,~~~
T = \prod_{j=1}^{2 N_c-2l} (v^2-q_j^2) .
\label{soopr1}
\eeq
Under the assumption concerning the projected curve on $v,t$ space
and the form of $w$,
we find 
\beq
\label{soowdef}
w  =  N + \frac{H'}{S} (t-C),
\eeq
where $N$ and $H'$ are some polynomials in $v$.
Remember that the fivebrane configuration has to be 
invariant under the reflection
$w \rightarrow -w,v \rightarrow -v, t \rightarrow -t$ \cite{LaLoLo},
we see that $C,S$ and $T$ transform 
to $-C,-S$ and $T$ under this reflection respectively.
Thus we require that $H \equiv H'/v =H(v^2)$, and $N$ transforms to $-N$.
The boundary conditions (\ref{soocond})
completely fix $N$ as (\ref{soau9}) and 
the behaver of $w$ in the asymptotic region $t \sim v^{2N_c-1}$ 
as (\ref{soau99}).
Note that $w$ satisfies (\ref{soau12}) in the $SO(2 N_c+1)$ case also.


We now consider the $SO(2 N_c+1)$ pure Yang-Mills theory.
From the first boundary condition in (\ref{soocond})
we find
\beq
w = 2 v H(v^2) \frac{B_{2 N_c}(v^2)}
{\prod_{i=1}^{l}(v^2-p_i^2) }
 + {\cal O}(v^{-1}) =  \sum_{k=1}^{2 N_c} \mu_{2 k} v^{2 k-1}
+ {\cal O}(v^{-1}),
 \label{sooeq}
\eeq
which determines $H(v^2)$ in terms of $\mu_{2 k}$. 
The equation (\ref{sooeq}) agrees with 
the field theory equation (\ref{sooau30}).
In the case of $SO(2 N_c+1)$ with $N_f$ flavors,
we can also compute the dyon vevs 
along the singular locus as in the previous calculations.
At the $r^*$-th Higgs branch
the curves (\ref{soorotb}) 
remain two rational curves after the rotation as
\beq
\widetilde{C}_L\, \left\{ \begin{array}{l} t=2 v^{2 N_c-1} \\
w= 0 \end{array} \right.
\qquad
C_R\, \left\{ \begin{array}{l}
t=\frac{1}{2} \La_{N=2}^{2(2 N_c-N_f-1)} v^{2(N_f-N_c)+1}\\
w=\sum_{k=1}^{N_c} \mu_{2 k} v^{2 k-1} .
\end{array}
\right.
\label{soorotb1}
\eeq
Thus from the brane picture we can observe that the $r^*$-th Higgs branch
root is not lifted for arbitrary values of the parameters $\mu_{2 k}$.


We will compute the finite values of $w$ at $t=0, v^2=m_i^2$ which 
correspond to the eigen values of the meson vev and
compare these to the field theory results for the case with one massless dyon.
Assuming $N_f<N_c-1$, 
we find 
\beq
w_{(i)} = \pm \frac{1}{2} \tilde{\mu}_{2 N_c} \La_{N=2}^{2 N_c-2-N_f} p
m_i \frac{\prod_{j=1}^{N_f} (p^2-m_i^2)^{\frac{1}{2}}}{m_i^2-p^2}, ~~~
p^2=\frac{\mu_{2 (N_c-1)}}{\mu_{2 N_c}},
\label{sooWi}
\eeq
which agrees with (\ref{soobb8}) up to a factor $\sqrt{2}$ .
Here we change the perturbation $\Delta W$ and the boundary conditions (\ref{soocond})
as in the case of $Sp(2 N_c)$.

The meson vev for the case with more than
one massless dyon is calculated as 
\beqa
w_{(i)} &= & \pm \La_{N=2}^{2 N_c-1-N_f} m_i 
\sum_{j=1}^{l} \frac{p_j H(p_j^2)}{\prod_{k \neq j} (p_j^2-p_k^2)} 
\frac{\prod_{f=1}^{N_f} (p_j^2-m_f^2)^{\frac{1}{2}}}{p_j^2-m_j^2} \CR
& =& \mp \frac{\pa}{\pa m_i} \pare{ \La_{N=2}^{2 N_c-2-N_f} \sum_{j=1}^{l} 
\sq{\det(p_j-m)} w_j p_j }.
\eeqa

\section{Conclusions}

We have obtained the descriptions of the moduli space of vacua of 
the four dimensional $N=1$ and $N=2$ supersymmetric gauge theories with 
the gauge groups $Sp(2 N_c)$, $SO(2 N_c)$ and $SO(2 N_c +1)$
using the M theory fivebrane and the orientifold plane.

First of all,
we have constructed the fivebrane configuration
corresponding to the Higgs branches of the $N=2$ supersymmetric gauge theories,
especially, the $r^*$-th Higgs branch root
which is analogous to the baryonic branch root of $SU(N_c)$ theory.
From the field theory analysis, it is known
that this root is not lifted 
after the adjoint mass perturbation to break $N=2$ to $N=1$ supersymmetry.
This phenomenon has been verified from the explicit construction of  
the corresponding M theory fivebrane configuration.
It is important that in M theory we should view 
the orientifold plane as the deformation of the 
fivebrane induced by the presence of the RR charges.

We have also studied the monopole condensations and 
the meson vev in the $N=1$ supersymmetric gauge theories
by considering the rotated NS 5-branes configurations.
The results agree with the field theory results
for the vacua in the phase with a single confined photon.

In the fivebrane framework, 
$N=1$ gauge theories with the Landau-Ginzburg type superpotential 
\cite{Ka} \cite{KiTeYa} \cite{GiPeRa}
and the non trivial fixed points of the $N=1$
supersymmetric gauge theories 
\cite{ArDo} \cite{TeYa} \cite{TeYa2} are also studied
along the lines of \cite{DeOz} .
It will be interesting to study these issues for the 
gauge groups $Sp(2 N_c)$ and $SO(N)$.
 

\vskip10mm

\noindent
{\bf Acknowledgements}

I would like to thank my colleagues 
for useful advice and 
interesting conversations. 
I am especially grateful to S.-K. Yang for valuable discussions and 
carefully reading the manuscript.
It is also my pleasure to thank K. Hori for his helpful comments.
This work is supported by JSPS Research Fellowship for Young 
Scientists.  \\

\noindent
{\bf Note added}: 

As this article was being completed, we received the preprints
\cite{AhOhTa3} \cite{Ah}
which overlap parts of the present work.
However the boundary conditions (\ref{cond}), (\ref{socond}) and (\ref{soocond})
for the M theory fivebrane we impose are different from those 
in \cite{AhOhTa3} \cite{Ah}.
 
\newpage


\end{document}